\documentclass[prd,superscriptaddress,showpacs,nofootinbib,amsmath,amssymb,aps,10pt]{revtex4}

\usepackage{bm}
\usepackage{amsfonts}
\usepackage{latexsym}
\usepackage[latin1]{inputenc}
\usepackage{graphicx}
\usepackage{amsmath}
\usepackage{palatino}
\usepackage{mathpazo}
\usepackage{booktabs}
\usepackage{dcolumn}

\def\jnl@style{\it}
\def\aaref@jnl#1{{\jnl@style#1}}

\def\aaref@jnl#1{{\jnl@style#1}}

\def\aj{\aaref@jnl{AJ}}                   
\def\apj{\aaref@jnl{ApJ}}                 
\def\apjl{\aaref@jnl{ApJ}}                
\def\apjs{\aaref@jnl{ApJS}}               
\def\apss{\aaref@jnl{Ap\&SS}}             
\def\aap{\aaref@jnl{A\&A}}                
\def\aapr{\aaref@jnl{A\&A~Rev.}}          
\def\aaps{\aaref@jnl{A\&AS}}              
\def\mnras{\aaref@jnl{Mon.~Not.~Roy.~Astron.~Soc.}}             
\def\prd{\aaref@jnl{Phys.~Rev.~D}}        
\def\prc{\aaref@jnl{Phys.~Rev.~C}}	
\def\prl{\aaref@jnl{Phys.~Rev.~Lett.}}    
\def\qjras{\aaref@jnl{QJRAS}}             
\def\skytel{\aaref@jnl{S\&T}}             
\def\ssr{\aaref@jnl{Space~Sci.~Rev.}}     
\def\zap{\aaref@jnl{ZAp}}                 
\def\nat{\aaref@jnl{Nature}}              
\def\aplett{\aaref@jnl{Astrophys.~Lett.}} 
\def\apspr{\aaref@jnl{Astrophys.~Space~Phys.~Res.}} 
\def\physrep{\aaref@jnl{Phys.~Rep.}}      
\def\physscr{\aaref@jnl{Phys.~Scr}}       
\def\commat{\aaref@jnl{Comm.~Math.~Phys.}}              
\def\science{\aaref@jnl{Science}}               
\def\cqg{\aaref@jnl{Classical Quant.~Grav.}}            
\def\jpcs{\aaref@jnl{JPCS}}                                     
\def\ijmpd{\aaref@jnl{Int.~J.~Mod.~Phys.~D}}                    
\def\grg{\aaref@jnl{Gen.~Relat.~Gravit.}}               
\def\rpp{\aaref@jnl{Rep.~Prog.~Phys.}}          
\def\npa{\aaref@jnl{Nucl.~Phys.~A}}        
\def\lrr{\aaref@jnl{Living Rev.~Rel.}}                   
\def\jcap{\aaref@jnl{J.~Cosmology Astropart.~Phys.}} 	
\def\rmp{\aaref@jnl{Rev.~Mod.~Phys.}} 	

\allowdisplaybreaks[1]

\begin{document}

\title{Gravitational wave asteroseismology of fast rotating neutron stars with realistic equations of state}

\author{Daniela D. Doneva}
\email{daniela.doneva@uni-tuebingen.de}
\affiliation{Theoretical Astrophysics, Eberhard Karls University of T\"ubingen, T\"ubingen 72076, Germany}
\affiliation{INRNE - Bulgarian Academy of Sciences, 1784  Sofia, Bulgaria}

\author{Erich Gaertig}
\affiliation{Theoretical Astrophysics, Eberhard Karls University of T\"ubingen, T\"ubingen 72076, Germany}

\author{Kostas D. Kokkotas}
\affiliation{Theoretical Astrophysics, Eberhard Karls University of T\"ubingen, T\"ubingen 72076, Germany}
\affiliation{Department of Physics, Aristotle University of Thessaloniki, Thessaloniki 54124, Greece}

\author{Christian Kr\"uger}
\affiliation{School of Mathematics, University of Southampton, Southampton SO17 1BJ, United Kingdom}

\date{\today}

\begin{abstract}
In the present paper we study the oscillations of fast rotating neutron stars with realistic equations of state (EoS) within the Cowling approximation. We derive improved empirical relations for gravitational wave asteroseismology with $f$-modes and for the first time we consider not only quadrupolar oscillations but also modes with higher spherical order ($l=|m|=3,4$). After performing a systematic comparison with polytropic EoS, it is shown that the empirical relations found in this case approximately also hold for realistic EoS. Even more, we show that these relations will not change significantly even if the Cowling approximation is dropped and the full general relativistic case is considered, although the normalization used here (frequencies and damping times in the nonrotating limit) could differ considerably. We also address the inverse problem, i.e. we investigate in detail what kind of observational data is required in order to determine characteristical neutron star parameters. It is shown that masses, radii and rotation rates can be estimated quite accurately using the derived asteroseismology relations. We also compute the instability window for certain models, i.e. the limiting curve in a $T-\Omega$--plane where the secular Chandrasekhar-Friedman-Schutz (CFS) instability overcomes dissipative effects, and show that some of the modern realistic EoS will lead to a larger instability window compared to all of the polytropic ones presented so far in the literature. Additionally, we calculate the $r$-mode instability window and compare it with the $f$-mode--case. The overall results for the instability window suggest that it is vital to take into account oscillations with $l=3,4$ when considering gravitational wave asteroseismology using the $f$-mode in rapidly rotating neutron stars, as these modes can become CFS unstable for a much larger range of parameters than pure quadrupolar oscillations.
\end{abstract}


\maketitle

\section{Introduction}
The problem of studying neutron star oscillations has been considered for several decades now \cite{Kokkotas99a,Andersson03,Stergioulas03} and the current advance in gravitational wave detectors might lead to actual observations of these oscillations in the near future. Many scenarios for the excitation of such oscillations were suggested \cite{Andersson10}, one of the most promising is the formation of a proto-neutron star shortly after a core-collapse supernova. It is expected that in this stage of the neutron star evolution, various modes will be excited and some of them might produce detectable amounts of gravitational radiation, especially if they are unstable.

One of the major challenges following the detection of gravitational waves from unstable neutron stars is to infer its characteristic parameters like mass, radius and rotation rate via the observed data. Extensive studies were performed in this direction which examine the possible information one can obtain by observing one or several oscillation modes \cite{Andersson98a,Kokkotas01,Benhar04,Gaertig10}. Empirical relations for gravitational wave asteroseismology were presented there which relate the oscillation frequencies and damping times of different modes with characteristic properties of the star. But while the nonrotating case was extensively studied in full general relativity and with various realistic equations of state, the rotating case is still not fully examined yet \cite{Gaertig10}. This is the astrophysically relevant case since newborn neutron stars, as a major source of gravitational waves, are supposed to be rapidly rotating and proper asteroseismology has to take this into account. Furthermore, fast rotating neutron stars can be destabilized by the so-called Chandrasekhar-Friedman-Schutz (CFS) instability \cite{Chandrasekhar70,Friedman78}, i.e. certain nonaxisymmetric modes can become unstable due to the emission of gravitational radiation for very high rotation rates of the star. If the requirements for this instability are met, the amplitude of the modes will grow exponentially even if they are only weakly excited.

A primary reason why most of the neutron star oscillation studies up to now were only considering the nonrotating case is the difficult numerical task of calculating the oscillation frequencies and damping times of fast rotating neutron stars. Also, a linear approach to full general relativity is usually employed as solving the full nonlinear problem for rotating neutron star oscillations is extremely demanding even on current high-performance hardware; only recently it became possible to address this problem properly \cite{Zink10}.\footnote{But even in this case, presently it is nearly impossible to use the nonlinear approach for a proper study of an extensive parameter space both in equations of state and stellar parameters.}

Up to now there is no numerical implementation available that solves the full general relativistic (GR) problem for oscillations of fast rotating neutron stars on a linear level and therefore, certain approximation have been implemented. An example is the so-called slow rotation approximation which was extensively used for studying the $r$-modes \cite{Kojima93,Andersson01,Ruoff01}, but is not suitable for studying the $f$-modes because the latter get secularly unstable at much higher rotation rates than the $r$-modes. Another commonly used approximation, more suitable for our goals, is the Cowling approximation where the perturbations of the metric are neglected and only the fluid variable perturbations are considered. Solving the linearized perturbation equations in the Cowling approximation was done for the first time in \cite{Gaertig08,Gaertig09} where the oscillation frequencies of the $f$, $g$ and $r$-modes for rapidly rotating polytropes were computed. These studies were extended later to also include differential rotation \cite{Kruger10,KrugerMaster}. An extensive parameter study concerning oscillation frequencies and damping times for fast rotating polytropes was performed in \cite{Gaertig10}, where several empirical relations for the $l=|m|=2$ $f$-mode have been derived and used for gravitational wave asteroseismology. The instability window and the evolution of neutron stars through this window were presented in \cite{Gaertig11,Passamonti12} for some selected polytropic models which favour the onset of the CFS-instability. Recently, a code which computes non-axisymmetric eigenmodes of rapidly rotating relativistic stars was developed in \cite{Yoshida12} by adopting the spatially conformally-flat approximation of general relativity.

In the present paper we will extend these results as well as the asteroseismology relations presented in \cite{Gaertig10} in two directions. First, we consider for the first time models with realistic equation of state (EoS).\footnote{We should note that possible deviations from the standard oscillation frequencies in GR could also arise when considering alternative models of neutron stars or alternatives theories of gravity \cite{Doneva12,Yazadjiev12,Sotani04,Sotani09} and this might eventually affect gravitational wave asteroseismology as well.} Although this extension seems logical and quite straightforward, it can lead to severe instabilities in numerical simulations when evolving the time dependent perturbation equations. One reason for this behaviour is that realistic EoS are usually given in a tabulated form where different equations of state are used to describe the nuclear matter at different densities. For example, a common practice is to use one EoS for the core and one or two others for the crust.This means that the $p(\rho)$ dependence is usually not a  smooth function.
Besides, there are also certain physical effects like sharp drops of the density profile in neutron stars and as a consequence in the fluid sound speed near the neutron drip point, which can cause numerical instabilities of the time evolution code and requires further adjustments. Also, a much higher numerical resolution should be used for realistic EoS when compared to polytropes due to these numerical difficulties.

The second important extension of the results in \cite{Gaertig10} refers to the empirical relations for gravitational wave asteroseismology presented there for the quadrupolar case. As already pointed out in \cite{Gaertig11,Passamonti12}, the $l=m=3$ and $l=m=4$ modes are much more promising to develop the CFS instability especially at lower rotation rates and consequently to produce observable amounts of gravitation radiation. Therefore, in the present work we will also derive the empirical relations for $l=|m|=3$ and $l=|m|=4$ $f$-modes which later can be used for gravitational wave asteroseismology.

We will also address the inverse problem -- given some potentially observed frequencies and damping times of a single neutron star model, one can use the empirical asteroseismology relations derived in this work to determine its mass, radius and rotation rate. This is the first study that also considers to solve this problem by using $f$-modes with $l>2$ as well. The results show that observing at least two $f$-modes with different spherical mode number $l$ can be used to determine the mass, radius and rotation rate of the star to a good accuracy.

In order to complete the study of the oscillations of fast rotating neutron stars with realistic EoS in the Cowling approximation, we also study the $f$-mode instability window for some of the realistic EoS. The results show that their instability window can be larger than the corresponding window for polytropic EoS considered in \cite{Gaertig11,Passamonti12}. This illustrates that some of the modern realistic equations of state are more favourable to the secular CFS-instability and could potentially lead to observable gravitational radiation signals from oscillating neutron stars. Finally, at the end we briefly compare the $f$-mode and the $r$-mode instability window for the considered sequences of rotating configurations.

This Paper is organized as follows: In Section \ref{sec:BasicRel} we comment on the formulation of the problem and the basic relations we are going to use. The extraction of the oscillation frequencies and damping times is considered in Section \ref{sec:DampingTimeFormulas}. The results for the computed equilibrium sequences and the asteroseismology relations are presented in Section \ref{sec:Asteroseism}. The inverse problem and its solution is addressed in Section \ref{sec:InverseProblem} and the instability window for some of the more optimistic candidates for the CFS-instability is computed in Section \ref{sec:InstabWindow}. We conclude this work with a summary and outlook.

\section{Basic relations} \label{sec:BasicRel}

The numerical implementation of the time evolution algorithm used in this work is mainly based on the experience gained in  \cite{Kruger10,KrugerMaster}, where the perturbation equations are set up in a formulation introduced by Vavoulidis and Kokkotas in \cite{Vavoulidis05,Vavoulidis07}. The independent variables which are evolved in time are not the primitive hydrodynamic quantities like velocity- or pressure-variations. Instead, the perturbations of the energy-momentum tensor are directly integrated and the only hydrodynamic quantity that enters the evolution equations explicitly is the speed of sound in the fluid. Additionally, we are also adopting an inertial frame of reference as opposed to a comoving frame utilized in \cite{Gaertig10,Gaertig08,Gaertig09}.

Here, we will give a brief introduction to the formulation that is used; more details can be found in the aforementioned literature. In spherical coordinates, the line element of the stationary and axisymmetric spacetime induced by a rotating neutron star takes the form
\begin{equation}\label{eq:metric}
ds^2 = -e^{2\nu} dt^2 + e^{2\psi}r^2 \sin^2 \theta (d\phi - \varpi dt)^2 + e^{2\mu}(dr^2 + r^2 d\theta^2)\;,
\end{equation}
where $\nu$, $\psi$ and $\varpi$ are functions of $r$ and $\theta$. Since we are working in the Cowling approximation \cite{Cowling41, McDermott83}, where perturbations of the metric are neglected and only fluid perturbations are considered, this is also the line element of the oscillating neutron star; i.e. no spacetime evolution is required. This approximation leads to good results for $g$-modes and higher-order $p$-modes while the error introduced for the fundamental $f$-mode can be as large as $30\%$ depending on the model, and decreases as $l$ is increased \cite{Finn88, Lindblom90, Yoshida97}.

The perturbation of the energy-momentum tensor in linearized GR can be written as
\begin{equation}
\delta T^{\mu\nu} = (\delta \epsilon + \delta p) u^\mu u^\nu + (\epsilon + p) (\delta u^\mu u^\nu + u^\mu \delta u^\nu) + \delta p  g^{\mu\nu}\;,
\label{pertEnergyMomentum}
\end{equation}
where $\epsilon$ is the energy density of the star, $p$ is the pressure, $u^\mu$ is the fluid four velocity, $g^{\mu\nu}$ is the metric tensor and $\delta (..)$ denotes the perturbation of the corresponding quantity. The four velocity can be represented as $u^\mu = (u^t,0,0,u^\phi)$ in spherical coordinates and the angular velocity $\Omega$ of the star is defined by
\begin{equation}
\Omega = \frac{u^\phi}{u^t}\;.
\end{equation}
 The perturbations of the energy-momentum tensor \eqref{pertEnergyMomentum} can also be written in the following matrix form
\begin{equation}
\delta T^{\mu\nu} = \left( \begin{array}{cccc} Q_1 & Q_3 & Q_4 & Q_2 \\ Q_3 & Q_6 & 0 & \Omega Q_3 \\
                                                Q_4 & 0 & Q_6/r^2 & \Omega Q_4 \\ Q_2 & \Omega Q_3 & \Omega Q_4 & Q_5 \end{array} \right)\;,
\end{equation}
where $Q_1,\ldots,Q_4$ are given by
\begin{eqnarray}\label{eq:Q_variables}
&& Q_1 = (\delta \epsilon + \delta p) (u^t)^2 + 2 (\epsilon + p) u^t \delta u^t + \delta p g^{\mu\nu}, \notag \\
&& Q_2 = (\delta \epsilon + \delta p) (u^t)^2 \Omega + (\epsilon + p) (\delta u^\phi + \Omega \delta u^t) u^t + \delta p g^{t\phi}, \\
&& Q_3 = (\epsilon + p) u^t \delta u^r,  \notag\\
&& Q_4 = (\epsilon + p) u^t \delta u^\theta\;,  \notag
\end{eqnarray}
and $Q_5$ and $Q_6$ can be expressed as combinations of $Q_1$ and $Q_2$.

The evolution equations for the $Q_i$-variables are derived from the conservation law of energy momentum. Thus in the Cowling approximation we have
\begin{equation} \label{eq:Pert_EM_Tensor}
\nabla_\nu (\delta T^{\mu\nu}) = 0\;,
\end{equation}
where $\nabla_\nu$ is the covariant derivative with respect to the metric (\ref{eq:metric}). This relation provides us with four evolution equations for the quantities $Q_1,..,Q_4$. The perturbations of the primitive fluid variables can then be reconstructed from the $Q_i$-variables by inverting relations \eqref{eq:Q_variables} in combination with two additional relations. First, because the fluid four-velocity is normalized, we obtain
\begin{equation}
\delta u^{t} = -\frac{u_\phi}{u_t} \delta u^\phi\;,
\end{equation}
which shows that only three out of the four components of $\delta u^\mu$ are independent. Second, the perturbations we consider are typically adiabatic and thus $\delta p$ and $\delta \epsilon$ are connected by
\begin{equation}
\delta p = c^2_s \delta \epsilon\;,
\end{equation}
where $c_s$ is the speed of sound in the fluid.

After time evolving the relevant equations with appropriate boundary conditions, we obtain the frequencies and eigenfunctions of the oscillation modes by post-processing the simulation data. When associating corresponding values of $l$ and $m$ to them, we have to keep in mind the following: In the nonrotating case, the fundamental mode frequencies only depend on the spherical mode index $l$, i.e. there is a degeneracy in the index $m$ so that frequencies for fixed values of $l$ but different values of the azimuthal mode number $m$ are identical. When rotation comes into play the picture changes considerably -- the degeneracy in $m$ is removed and a nonrotating mode with a certain spherical index $l$ splits into $(2l+1)$ modes with different frequencies\footnote{$m$ ranges from $-l$ to $+l$.}.
Also in the rotating case, the angular part of the modes cannot be represented by spherical harmonics any more, so strictly speaking it is not possible to associate a certain index $l$ to the oscillation. Instead, the value of $l$ is defined as the corresponding value that this particular mode would exhibit in the nonrotating limit.

\section{Extraction of oscillation frequencies and damping times} \label{sec:DampingTimeFormulas}

In order to obtain empirical relations which can later be used for asteroseismology and for investigating the instability window of fast rotating neutron stars with realistic EoS, we have to extract both the oscillation frequencies and the damping times of the modes. Extracting the frequencies is straightforward -- one only has to perform a Fourier transform on the computed time series. As explained above, the use of realistic equations of state causes some numerical instabilities. For testing purposes, we first compared the oscillation frequencies for nonrotating models obtained with our time evolution code with  a 1D-code that solves the time independent perturbation equations in the Cowling approximation
\footnote{The 1D code is supposed to be more stable.} \cite{Yazadjiev12,Doneva12}. The frequencies computed with both codes are in good agreement.

Calculating the damping times of modes is more involved since this quantity is defined as the inverse of the imaginary part of its complex frequency. But since we are working in the Cowling approximation, the gravitational radiation degrees of freedom are neglected and the oscillation frequencies are purely real. We have to use an alternative way of calculating the damping times and a common approach is to use an approximate Newtonian formula, where the emission of gravitational waves is related to the mutlipole moments of the neutron star \cite{Thorne80,Balbinski82,Balbinski85}. We will adopt this relation as it leads to satisfactory results also in the general relativistic case \cite{Balbinski85,Gaertig10}. Moreover it is expected that the deviation in the calculated damping times due to the Cowling approximation is much larger than the corresponding error introduced by using the approximate multipole formula.

In a more precise formulation, if one assumes that the time dependence of all perturbation variables is harmonic, i.e. $\sim e^{i\omega t}$ and the energy radiated per cycle is much smaller than the energy of the mode, the damping time can be estimated by \cite{Balbinski85,Lockitch98,Ipser91}
\begin{equation}
\label{eq:dampingTime}
\frac{1}{\tau} = -\frac{1}{2E}\frac{dE}{dt}\;,
\end{equation}
where $E$ is the energy of the mode in the comoving frame and $dE/dt$ is the energy loss.
The energy is given by
\begin{equation}
\label{eq:EnergyModes}
E = \frac{1}{2}\int\left[\rho\delta u^{a} \delta u^{\ast}_{a} + \left(\frac{\delta p}{\rho} + \delta \Phi\right)\delta\rho^{\ast} \right]d^3x\;,
\end{equation}
where $\rho$ is the rest-mass density, and $\delta\rho$, $\delta p$, $\delta u_{a} $, $\delta\Phi$ are the perturbations of the rest-mass density, the  pressure, the spatial part of the fluid velocity and the gravitational potential respectively.
Also, within the Cowling approximation we neglect the term proportional to  $\delta\Phi$.

The energy loss due to gravitational radiation can be computed using
\begin{equation}
\label{eq:dEdt_GW}
\frac{dE}{dt} = - \omega_i(\omega_i + m\Omega)\sum_{l \ge 2} N_l \omega_i^{2l}(|\delta D_{l}^m|^2 + |\delta J_{l}^m|^2)\;,
\end{equation}
where $\omega_i$ is the frequency of the mode in the inertial frame. We will later use also the corresponding frequency $\omega_c$ in a comoving frame and both of them are related via the standard relation
\begin{equation}\label{eq:SigmaInertialComoving_Relation}
\omega_c = \omega_i + m\Omega\;.
\end{equation}
The quantities $\delta D_{lm}$ and  $\delta J_{lm}$ are the mass and the current multipole moments of the perturbation given by
\begin{eqnarray}
 &&\delta D_{lm} = \int \delta \rho \, r^l Y^{m\,\ast}_l d^3x \label{eq:Dlm} \\
 &&\delta J_{lm} = 2 \sqrt{\frac{l}{l+1}} \int r^l (\rho \, \delta u_a + \delta \rho \, u_a) Y^{a,B\, \ast}_{lm} \label{eq:Jlm}\;,
\end{eqnarray}
where the $Y^{m}_l$ are the standard spherical harmonics and the $Y^{a,B}_{lm}$ are the magnetic type vector spherical harmonics \cite{Thorne80, Lockitch98}.

For the pressure modes ($p$-modes) in general and the $f$-mode in particular, the current multipole moments can be neglected as the mass multipoles represent the dominant contribution to the gravitational wave damping. For the $r$-modes on the other hand it is the opposite case -- the current multipoles account for the main contribution there.

In our sign convention, the nonaxisymmetric modes with $m<0$ are prograde and their frequencies in the inertial frame increase when increasing the rotation rate as can be seen from equation \eqref{eq:SigmaInertialComoving_Relation}. Oscillations with an azimuthal index  $m>0$ on the other hand are retrograde and their inertial frame frequencies decrease while increasing the stellar rotation. For fast rotating stars these frequencies can reach negative values, effectively turning them into prograde modes with respect to an observer in the inertial frame, and this turning point marks the onset of the Chandrasekhar-Friedman-Schutz (CFS) instability \cite{Chandrasekhar70,Friedman78}. The basic essence of this instability is that retrograde modes in the comoving frame are dragged forward by rotation, thereby becoming prograde in the inertial frame and getting secularly unstable due to the emission of gravitational waves. This can also be seen in equation \eqref{eq:dEdt_GW} -- for negative $\omega_i$ and positive $\omega_c$ the energy of the mode increases with time. In the limiting case of $\omega_i = 0$, i.e. at the onset of the instability, a neutral oscillation appears that exhibits a stationary mode pattern in the inertial frame and does not emit gravitational radiation at all. As relation \eqref{eq:SigmaInertialComoving_Relation} shows, modes with higher azimuthal index $m$ can potentially reach negative mode frequencies at smaller rotation rates, therefore favouring the CFS-instability. That is the reason why in this study we will focus on nonaxisymmetric oscillations with the largest allowed value of $m$ for a given $l$, i.e. when $|m|=l$.

As one can see from equations \eqref{eq:dEdt_GW}, \eqref{eq:SigmaInertialComoving_Relation} if one only takes into account the emission of gravitational waves, for every rotation rate of the star one can choose modes with corresponding large values of $m$, so that $\omega_i < 0$ and they instantly get unstable. However, several dissipative effects acting on different timescales counteract the exponential growth of the CFS-instability and might lead to a saturation of the $f$-mode \cite{Kastaun10} or eventually suppress it completely. In practice it turns out that for the vast majority of realistic EoS it is relevant to consider only modes with $m < 5$.

If we assume that neutron star matter is a mixture of protons, neutrons and electrons, the dissipation is mainly due to the familiar shear and bulk viscosities. Taking into account these additional effects, the total damping time $\tau$ of a mode can be estimated by
\begin{equation} \label{eq:fullTau}
\frac{1}{\tau} = \frac{1}{\tau_\zeta} + \frac{1}{\tau_\eta} + \frac{1}{\tau_{GW}}\;,
\end{equation}
where $\tau_{GW}$ denotes the gravitational wave damping time, $\tau_{\zeta}$ is the bulk viscosity damping time and $\tau_{\eta}$ represents the shear viscosity damping time. If $\tau$ is negative then the mode is exponentially growing on this timescale, i.e. it is unstable \footnote{We should note that only $\tau_{GW}$ is negative for CFS unstable modes while both $\tau_\zeta$ and $\tau_\eta$ are always positive.}.

We already provided relations for the gravitational wave damping time, i.e. equations \eqref{eq:dampingTime}--\eqref{eq:dEdt_GW}. Shear and bulk viscosity timescales are computed by standard relations derived in Newtonian theory \cite{Ipser91,Lockitch98} and are given by

\begin{equation}
\frac{1}{\tau_{\zeta}} = -\frac{1}{2E}\int \zeta \delta \theta \, \delta \theta^* d^3 x\;,
\end{equation}

\begin{equation}
\frac{1}{\tau_{\eta}} = - \frac{1}{E} \int  \eta \delta \sigma^{ab} \sigma_{ab}^* d^3 x\;,
\end{equation}
where $\zeta$ and $\eta$ are the bulk and shear viscosity coefficients respectively. The shear $\delta \sigma_{ab}$ and the expansion $\delta \theta$ of the perturbations are given by
\begin{eqnarray}
&& \delta \theta = \nabla_c \delta u^c\;, \\ \notag \\
&& \delta \sigma_{ab} = \frac{1}{2} (\nabla_a \delta u_b + \nabla_b \delta u_a - \frac{2}{3} g_{ab} \nabla_c \delta u^c)\;, \notag
\end{eqnarray}
and we use the following values for the coefficients $\zeta$ and $\eta$
\begin{eqnarray}
&& \zeta = 6 \times 10^{25} \left( \frac{1{\rm Hz}}{\omega+m\Omega} \right)^2 \left( \frac{\rho}{10^{15} {\rm g}\,{\rm cm^{-3}}} \right)^2  \left( \frac{T}{10^{9} {\rm K}} \right)^6 {\rm g}\, {\rm cm}^{-1} {\rm s}^{-1}\;, \\ \notag \\
&& \eta = 2 \times 10^{18} \left( \frac{\rho}{10^{15} {\rm g}\,{\rm cm^{-3}}} \right)^{9/4}  \left( \frac{10^{9} {\rm K}}{T} \right)^2 {\rm g}\, {\rm cm}^{-1} {\rm s}^{-1}\;,
\end{eqnarray}
derived for a mixture of neutrons, protons and electrons in a normal state, i.e. without superfluid or superconducting components \cite{Sawyer89,Cutler87}. These formulas are strictly valid only in the linear regime. If the amplitude of the modes grows considerably, nonlinear effects will be present \cite{Alford12,Kastaun10} which also lead to additional damping mechanisms. In our study though, we only consider linear perturbations and the relations given above are fully applicable in this case.

\section{Asteroseismology}\label{sec:Asteroseism}
\subsection{The background models}
In our simulations we utilize the rns code developed by N. Stergioulas \cite{Stergioulas95,Nozawa98} to construct background models of rotating relativistic stars. It is particularly suitable for our goals since the code can deal with neutron stars rotating at arbitrarily high rotational frequencies (up to the mass-shedding limit) and it is also able to handle realistic equations of state.

In order to derive empirical relations for gravitational wave asteroseismology, we choose to study the oscillation spectrum for equilibrium configurations with fixed central energy density and different rotation rates ranging from zero up to the mass-shedding limit, i.e. the Kepler limit. It should be pointed out that during the evolution of a young neutron star, it is actually the baryon mass that remains fixed and not the central energy density. Therefore, our sequences do not correspond to the evolution (spin down) of a single star. But since our goal is to obtain empirical relations, the actual sequence of models is not important. All we need in this case is just a big pool of configurations with different masses, rotational frequencies and equations of state. The reason for adopting constant central energy density sequences is just for simplicity. Later in this work when the $f$-mode instability window is studied, the correct sequences for this case, i.e. models with constant baryon mass, are computed because at this point the evolution of a single neutron star has to be tracked.

\begin{figure}[ht!]
\centering
\includegraphics[width=0.55\textwidth]{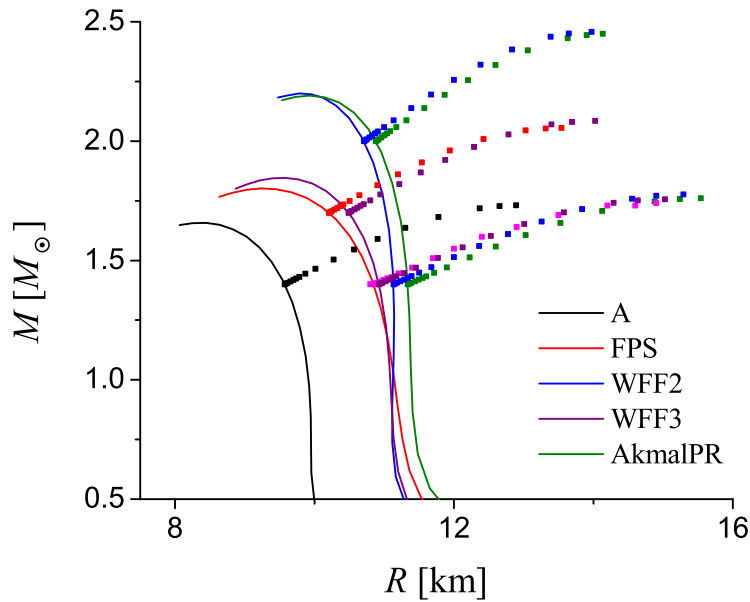}
\caption{The mass-radius relations for the background models listed in Table \ref{tbl:models}. The sequences branch off their nonrotating counterparts at the corresponding equilibrium curve (solid lines) and increase in masses and radii as the rotation rate is increased (dotted lines). }
\label{Fig:M_R_AllEOS_RotBG}
\end{figure}

We will consider five equations of state which are listed in Table \ref{tbl:EOS List}. The corresponding mass-radius--relation for these EoS are depicted in Figure \ref{Fig:M_R_AllEOS_RotBG} (solid lines). Two of these EoS (WFF2 and AkmalPR) reach the two solar mass barrier and are in agreement with the current observational constraints \cite{Demorest10,Manousakis12,Antoniadis13,Lattimer12}, but we use a larger set of equations of state in order to show the robustness of our asteroseismology approximations and to better explore the parameter space.

\begin{table*}[ht!]
\begin{center}
\caption{The equations of state utilized in this study. }\label{tbl:EOS List}
\begin{tabular}{|l|p{14cm}|}
\hline
{\bf EoS} &  {\bf Description}\\
\hline
\hline
FPS & Equation of state by Lorenz, Ravenhall and Pethick \cite{FPS}. A modern version of the equation of state by Friedman and Pandharipande  \cite{Pandharipande81}.\\[0.1in]
WFF2 & Equation of state by Wiringa, Fiks and Fabrocini \cite{WFF}, denoted ``UV14+UVII'' in their paper. Matched to Negele and Vautherin \cite{Negele73} at low densities. \\[0.1in]
WFF3 & Equation of state by Wiringa, Fiks and Fabrocini \cite{WFF}, denoted ``UV14+TNI'' in their paper. Matched to the FPS equation of state at low densities.\\[0.1in]
A & Equation of state by Arnett and Bowers \cite{Pandharipande71,A}, denoted ``EOS A'' in their paper. \\[0.1in]
AkmalPR & Equation of state by Akmal, Pandharipande, and Ravenhall \cite{AkmalPR}. Matched with a SLY4 crust \cite{Douchin01}. \\[0.01in]
\hline
\end{tabular}
\end{center}
\end{table*}

For most of the EoS, two rotational sequences with different central energy densities are computed  -- the first sequence starts with a mass of $M=1.4 M_\odot$ in the nonrotating limit while the second one is put close to the maximum allowed mass for the corresponding EoS; Table \ref{tbl:models} summarizes the characteristic neutron star parameters for both sequences. Additionally, the mass-radius--relationships for the rotating configurations with constant central density are also depicted in Figure \ref{Fig:M_R_AllEOS_RotBG} (dotted lines). They start from the corresponding nonrotating models and reach up to the mass-shedding limit. In this Figure, a well know fact can also be observed -- if the central energy density is kept fixed, the mass and the radius of the neutron star increase with rotation due to the presence of the centrifugal force which supports pressure to sustain gravity.

\begin{table*}[]
\begin{center}
    \caption{The characteristic nonrotating neutron star parameters for the sequences used in this study.}\label{tbl:models}
\begin{tabular}{|p{5cm}p{2cm}p{2cm}p{2cm}|}
\hline
{\bf EoS} &  ${\bm\rho_{\bm c}\bm [ {\rm\bf g/cm^3}\bm ]}$   & $\bm M_0\bm [\bm M_{\bm\odot}\bm ]$  & $\bm R_0\bm [{\rm\bf km}\bm ]$     \\
\hline
\hline
 FPS  & $1.30 \times 10^{15}$   & $1.4$  &  $10.85$  \\[0.1in]
 FPS  & $2.02 \times 10^{15}$   &  $1.7$ & $10.21$   \\[0.1in]
 WFF2  & $1.04 \times 10^{15}$   & $1.4$  & $11.13$   \\[0.1in]
 WFF2  & $1.64 \times 10^{15}$   & $2.0$  &  $10.71$  \\[0.1in]
 WFF3  & $1.21 \times 10^{15}$   & $1.4$  & $10.92$  \\[0.1in]
 WFF3  & $1.75 \times 10^{15}$   &  $1.7$ & $10.49$   \\[0.1in]
 A  & $1.85 \times 10^{15}$    &  $1.4$ & $9.57$   \\[0.1in]
 AkmalPR  & $1.01 \times 10^{15}$   & $1.4$  & $11.34$   \\[0.1in]
 AkmalPR  & $1.60 \times 10^{15}$   & $2.0$  &  $10.88$  \\[0.01in]
\hline
\end{tabular}
\end{center}
\end{table*}

\subsection{Results}
In this Section, the results for oscillation frequencies and damping times of rotating neutron stars with realistic EoS for the $l=|m|=2, 3,4$ case are presented. We will restrict ourselves only to consider  the fundamental $f$-modes as they are one of the most promising candidates to develop a CFS-unstable phase and thus to emit considerable amounts of gravitational radiation. Similar asteroseismology relations were already obtained by Gaertig and Kokkotas in \cite{Gaertig10} for neutron stars with polytropic EoS in the $l=|m|=2$ case.

\subsubsection{Asteroseismology relations for oscillation frequencies}\label{asteroForFrequencies}

\begin{figure}[ht!]
\centering
\includegraphics[width=0.5\textwidth]{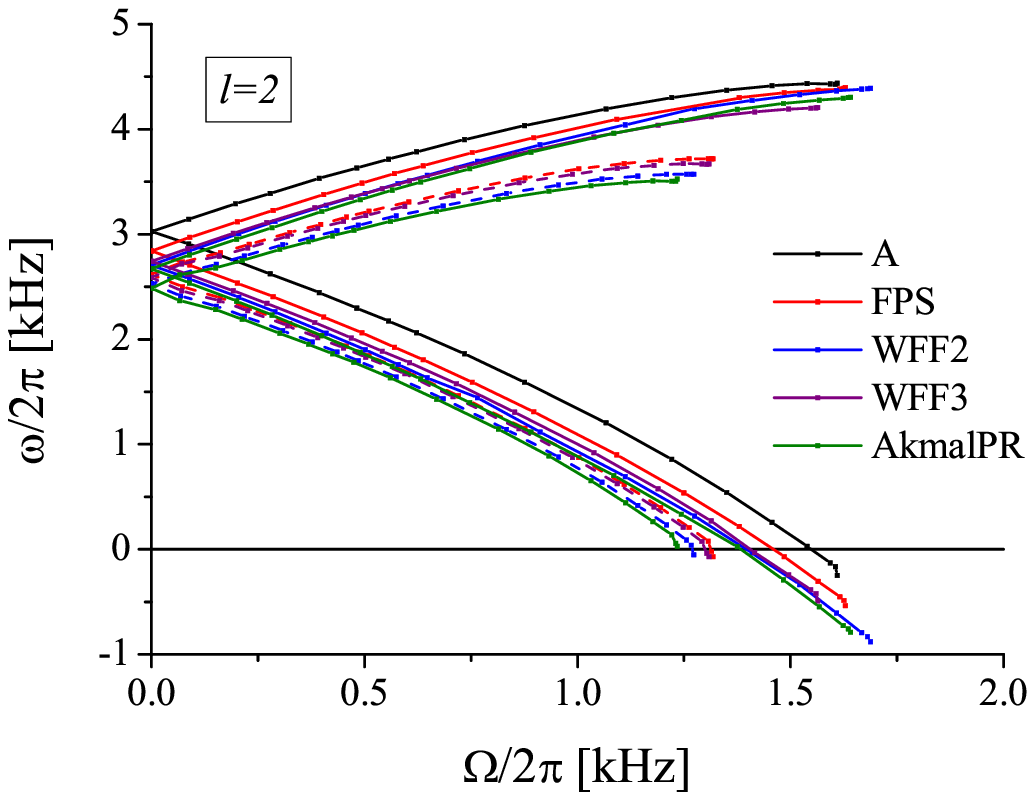}
\includegraphics[width=0.48\textwidth]{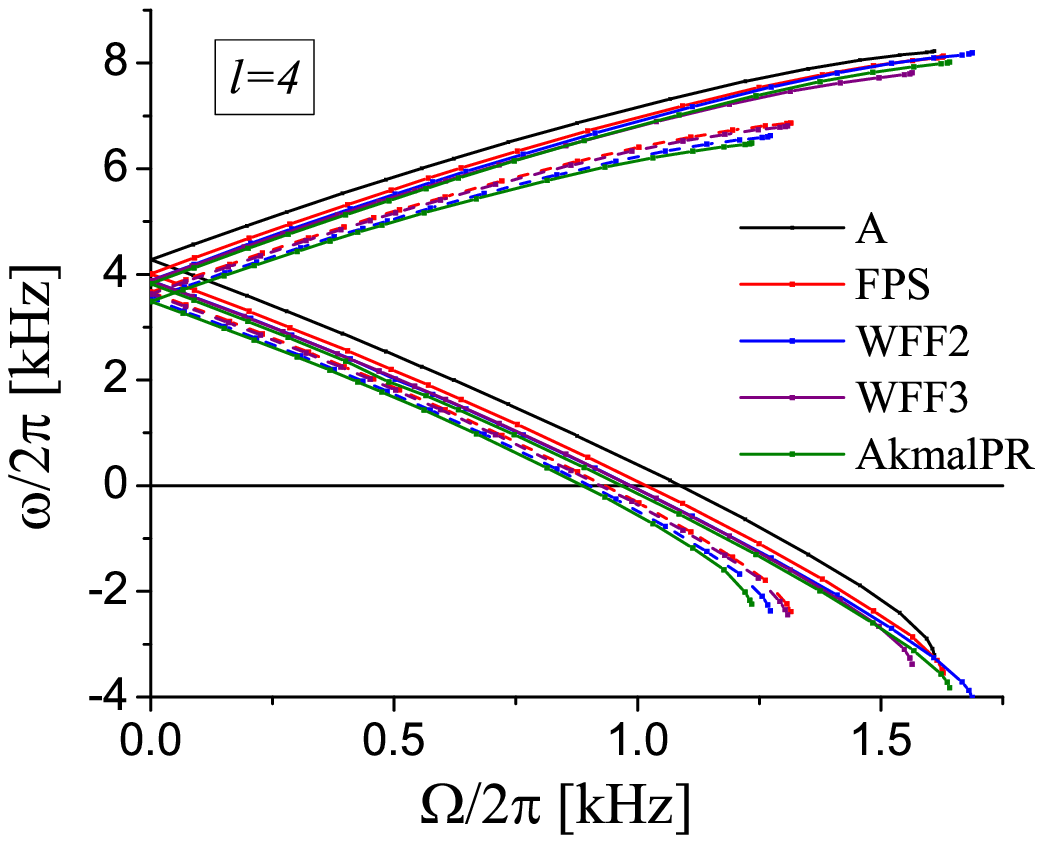}
\caption{$f$-mode frequencies in inertial frame corresponding to $l=|m|=2$ and $l=|m|=4$ as a function of the rotation rate for both co- and counterrotating branches. The dashed lines correspond to the less compact configurations in Table \ref{tbl:models}. } \label{Fig:sigmaInertial_l=2_4}
\end{figure}

The characteristic mode splitting of nonaxisymmetric modes in rotating neutron stars can be observed in Figure \ref{Fig:sigmaInertial_l=2_4} where the frequencies of the $l=|m|=2$ and $l=|m|=4$ $f$-modes are depicted for the sequences in Table \ref{tbl:models}. Each branch corresponds to a particular model with constant central energy density and ranges from $\Omega=0$ to $\Omega=\Omega_K$, where $\Omega_K$ is the Kepler frequency. The upper (stable) branches correspond to the corotating modes with $m=-l<0$ while the lower (potentially unstable) ones show the counterrotating modes with $m=l>0$. The Figure shows the well-known fact, that in the quadrupolar case some of the lower branches actually never reach the CFS unstable regime. Also, given a fixed EoS the oscillation modes of more compact models are able to reach negative frequencies at lower rotation rates. For higher values of the spherical index $l=m>2$, all lower branches become unstable after a certain critical rotation rate $\Omega_c$.

In order to do proper gravitational wave asteroseismology, one has to derive empirical relations that connect the observed oscillation frequencies to the neutron star properties in an EoS-independent way. When deriving such relations we will mainly stick to the approach taken in \cite{Gaertig10} in order to show differences and similarities between polytropic and realistic equations of state, and furthermore to consistently generalize the relations given there for the $l=|m|=3, 4$ case.

In the nonrotating case, the frequencies are roughly proportional to the square root of the mean density \cite{Andersson98a,Kokkotas01}. When rotation is added there is another parameter which has to be determined -- the angular velocity of the star. It turns out that it is convenient to use two independent relations in this case \cite{Gaertig10,Gaertig08}. The first one provides the normalized oscillation frequency as a function of the normalized rotation rate, where the relations are normalized by the frequency in the nonrotating limit and the Kepler frequency respectively. Naturally the second one correlates the frequencies in the nonrotating limit with the mean density of the star, similar to the relations obtained in \cite{Andersson98a,Kokkotas01}.

We will follow \cite{Gaertig10} and use the oscillation frequencies in the comoving frame. As it turned out, in this frame the spread of the frequencies for different EoS is considerably smaller than in the inertial frame \cite{Gaertig08}, therefore providing a natural frame for this model-independent fitting. Another nice feature of the comoving frame is that in contrast to the inertial frame, mode frequencies of both branches never get negative there.

\begin{figure}[ht!]
\centering
\includegraphics[width=0.6\textwidth]{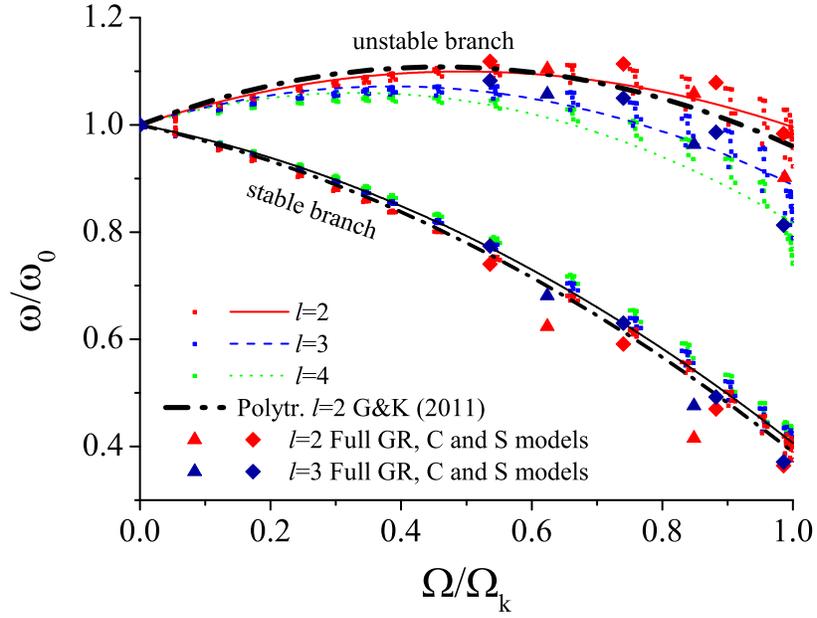}
\caption{The normalized oscillation frequencies as a function of the normalized rotation rate in the comoving frame. The results for $l=|m|=2, 3, 4$ and for  all of the configurations in Table \ref{tbl:models} are depicted. }
\label{Fig:sigmaComoving_l=234}
\end{figure}

The normalized frequencies in the comoving frame $\omega_c/\omega_0$ ($\omega_0$ is the frequency in the nonrotating limit) as a function of  $\Omega/\Omega_K$ for all the EoS considered in this work are shown in Figure \ref{Fig:sigmaComoving_l=234}. It should be noted here that in the comoving frame the order of the two branches is reversed, i.e. the potentially unstable branches attain larger frequencies than the stable ones in contrast to the depiction in the inertial frame.

The relations for different values of $l$, shown in Figure \ref{Fig:sigmaComoving_l=234}, can be fitted very accurately with a polynomial of second order. We thus obtain the following relations for the frequencies of the potentially unstable branches $\omega_{c}^u$,

\noindent
for $l=m=2$:
\begin{eqnarray} \label{eq:sigmaRotUnstFit_l2}
&&\frac{\omega_{c \mbox{ } l=2}^u}{\omega_0} = 1 + 0.402 \left(\frac{\Omega}{\Omega_K} \right) - 0.406 \left(\frac{\Omega}{\Omega_K} \right)^2\;,
\end{eqnarray}
for $l=m=3$:
\begin{eqnarray} \label{eq:sigmaRotUnstFit_l3}
&&\frac{\omega_{c \mbox{ } l=3}^u}{\omega_0} = 1 + 0.373 \left(\frac{\Omega}{\Omega_K} \right) - 0.485 \left(\frac{\Omega}{\Omega_K} \right)^2\;,
\end{eqnarray}
and for $l=m=4$
\begin{eqnarray} \label{eq:sigmaRotUnstFit_l4}
&&\frac{\omega_{c \mbox{ } l=4}^u}{\omega_0} = 1 + 0.360 \left(\frac{\Omega}{\Omega_K} \right) - 0.543 \left(\frac{\Omega}{\Omega_K} \right)^2\;.
\end{eqnarray}
As one can see from Figure \ref{Fig:sigmaComoving_l=234}, the frequencies for the stable branches $\omega_{c}^s$ can be fitted very well by a single quadratic polynomial for all values of $l$ and we obtain
\begin{eqnarray} \label{eq:sigmaRotStableFit_l234}
&&\frac{\omega_{c \mbox{ }}^s}{\omega_0} = 1 - 0.235 \left(\frac{\Omega}{\Omega_K} \right) - 0.358 \left(\frac{\Omega}{\Omega_K} \right)^2\;.
\end{eqnarray}

\begin{figure}[ht!]
\centering
\includegraphics[width=0.49\textwidth]{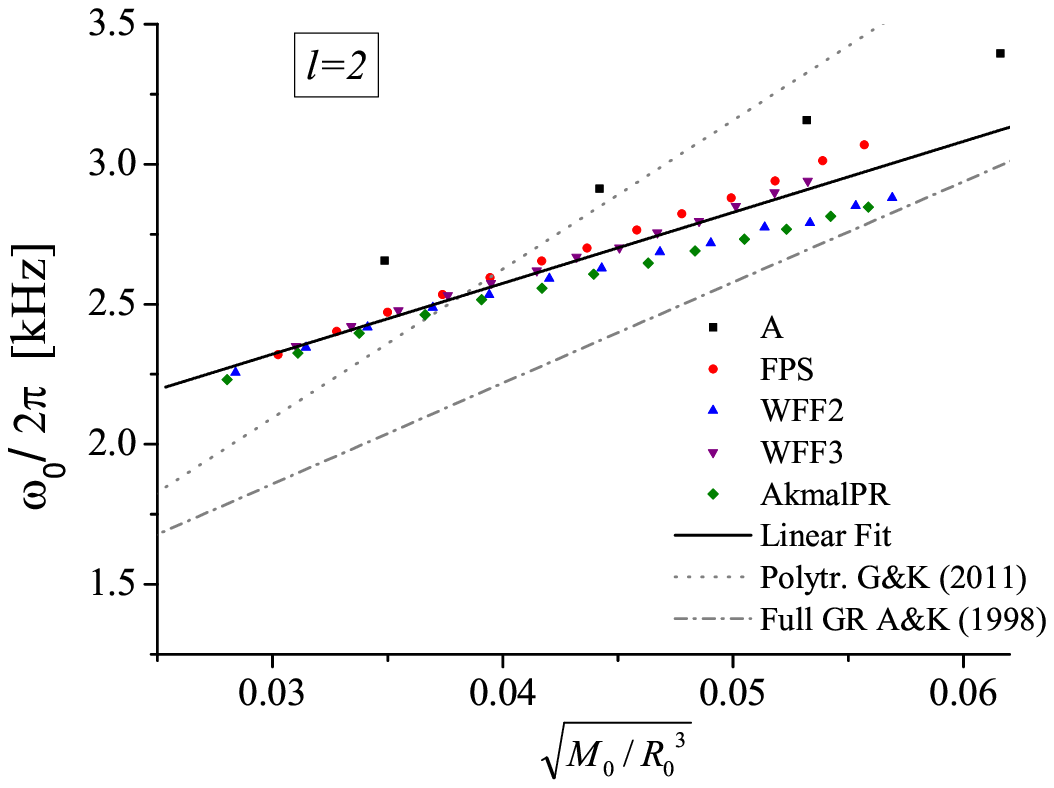}
\includegraphics[width=0.49\textwidth]{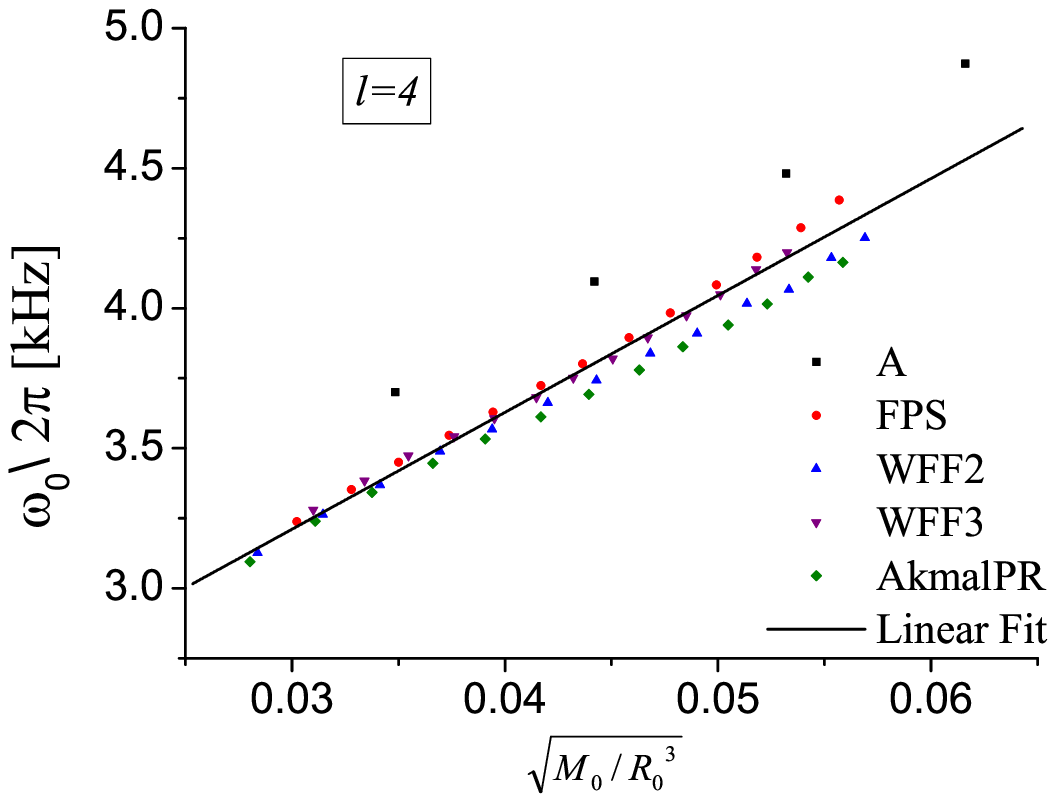}
\caption{Mode frequencies for $l=2, 4$ as a function of the average density in the nonrotating limit.} \label{Fig:StaticFreq}
\end{figure}

As discussed previously, the relations \eqref{eq:sigmaRotUnstFit_l2}--\eqref{eq:sigmaRotStableFit_l234} have to be supplemented with additional information on how the mode frequencies in the nonrotating limit $\omega_0$ depend on the neutron star mass and radius. It has been shown  \cite{Andersson98a,Kokkotas01} that the average density is a good measure to parametrize this dependency and Figure \ref{Fig:StaticFreq} shows the results with our pool of configurations.

By making a linear approximation similar to \cite{Andersson98a,Kokkotas01,Gaertig10}, the following relations are obtained,

\noindent
for $l=2$:
\begin{equation}
\label{eq:sigmaZeroNonrot2}
\frac{1}{2\pi}\omega_{0\mbox{ }l=2}\,\mathrm{[kHz]} = 1.562 +  1.151 \left(\frac{\bar M_0}{\bar R_0^3}\right)^{1/2}\;,
\end{equation}
for $l=3$:
\begin{equation}
\label{eq:sigmaZeroNonrot3}
\frac{1}{2\pi}\omega_{0\mbox{ }l=3}\,\mathrm{[kHz]} = 1.764 + 1.577 \left(\frac{\bar M_0}{\bar R_0^3}\right)^{1/2}\;,
\end{equation}
for $l=4$:
\begin{equation}
\label{eq:sigmaZeroNonrot4}
\frac{1}{2\pi}\omega_{0\mbox{ }l=4}\,\mathrm{[kHz]} = 1.958 + 1.898 \left(\frac{\bar M_0}{\bar R_0^3}\right)^{1/2}\;.
\end{equation}
Here we have introduced the dimensionless variables
\begin{equation}
\label{eq:M_and_R_Normalized}
\bar M = \frac{M}{\mathrm{1.4}\, M_\odot}\quad\mathrm{and}\quad\bar R = \frac{R}{\mathrm{10\,km}}\;.
\end{equation}
and the subscript $(..)_0$ indicates that these are the masses and radii of the nonrotating configurations.

In relations \eqref{eq:sigmaRotUnstFit_l2}--\eqref{eq:sigmaRotStableFit_l234}, the Kepler frequency $\Omega_K$  shows up as an additional free parameter. But $\Omega_K$ is roughly proportional to the average density as well, as it was shown in \cite{Friedman89,Haensel89,Lasota96,Stergioulas03}. Instead of using the relation given in these papers, we derive our own version obtained from fitting the data for the realistic EoS used here, which is more accurate for the considered range of masses, radii and EoS. We then obtain
\begin{equation} \label{eq:OmegaK}
\frac{1}{2\pi}\Omega_K {\rm [kHz]} = 1.716 \sqrt{\frac{\bar M_0}{{\bar R_0}^3}} - 0.189\;.
\end{equation}
This relation can be refined further by assuming that the coefficients are not constant but depend on the compactness $M/R$ \cite{Lasota96,Stergioulas03}. We prefer to use the relation in its current form, because it will prove to be useful later for the asteroseismology examples and additionally it also estimates the Kepler frequency with a very good accuracy -- for the models studied here the error is only up to approximately 2\%.

\begin{figure}[ht!]
\centering
\includegraphics[width=0.49\textwidth]{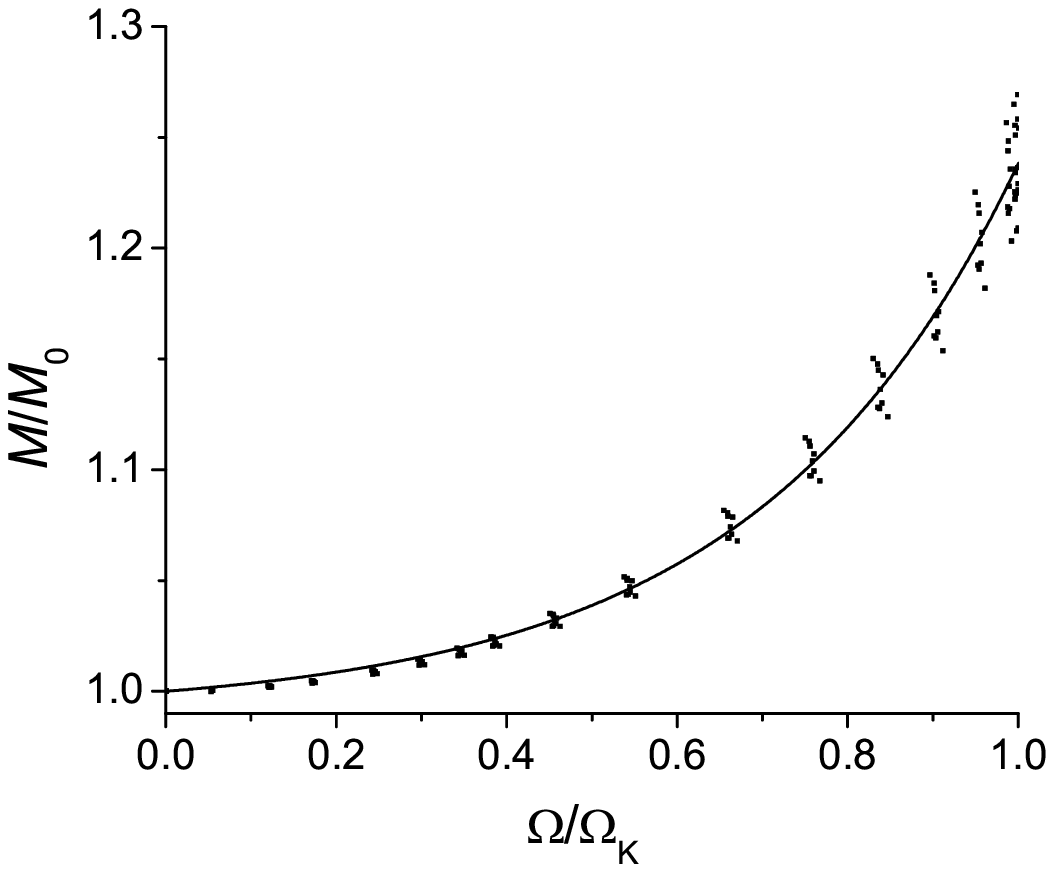}
\includegraphics[width=0.49\textwidth]{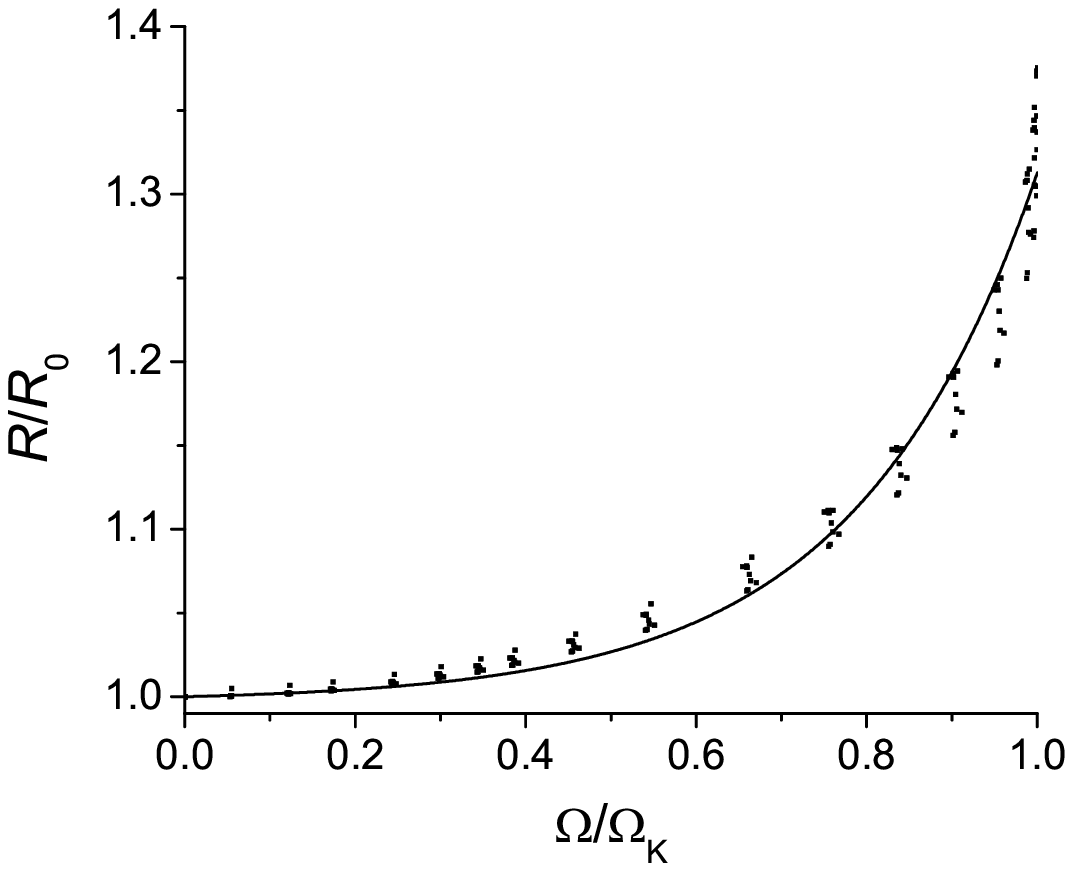}
\caption{The normalized mass (left panel) and radius (right panel) as a function of the normalized rotational frequency where $M_0$ and $R_0$ are the mass and the radius in the nonrotating limit.} \label{Fig:MRnorm}
\end{figure}

The last thing we have to specify in order to be able to use the above relations for gravitational wave asteroseismology is the following: The equations \eqref{eq:sigmaZeroNonrot2}--\eqref{eq:sigmaZeroNonrot4},\eqref{eq:OmegaK} are derived using nonrotating neutron star models. Therefore the masses and radii that enter in these equations are the masses and radii of the configurations in the nonrotating limit. As our goal is to be able to determine the parameters of the emitting rotating neutron stars we should know how masses and radii scale with rotation. We found out that it is convenient to derive an approximate relation for the normalized masses and radii as a function of $\Omega/\Omega_K$ and the results are plotted on Fig. \ref{Fig:MRnorm}. The data can be fitted well with an exponential function of the form $y = A + B\exp(Cx)$. Due to the normalization we have that $y(x)|_{x=0}=1$ which sets a constraint on the parameters of the fit, i.e. $A=1-B$. Thus we obtain the following relations for the normalized masses and radii

\begin{eqnarray}
&&\frac{M}{M_0} = 0.991 + 9.36 \times 10^{-3} \, \exp\left(3.28 \frac{\Omega}{\Omega_K}\right),  \label{eq:Mnorm_RotCorrection} \\  \notag \\
&&\frac{R}{R_0} = 0.997 + 2.77  \times 10^{-3} \, \exp\left(4.74 \frac{\Omega}{\Omega_K}\right). \label{eq:Rnorm_RotCorrection}
\end{eqnarray}
Using these relations we can obtain the mass and the radius of a rotating neutron star once we have determined the parameters in the nonrotating limit, $M_0$ and $R_0$, from equations \eqref{eq:sigmaZeroNonrot2}--\eqref{eq:sigmaZeroNonrot4},\eqref{eq:OmegaK}.

Let us draw a comparison with the polytropic case at this point. The thick black dash-dotted line in Figure \ref{Fig:sigmaComoving_l=234} represents the analytic relations found in \cite{Gaertig10} for polytropic EoS and quadrupolar modes ($l=2$). As one can see, the polynomial approximations of the stable branches in the case of polytropes and realistic EoS is quite similar. The corresponding fittings for the unstable branches are very similar as well and only for fast rotation rates one can see a certain divergence. In Figure \ref{Fig:sigmaComoving_l=234} we also plot the available results for mode frequencies in full general relativity, i.e. when the Cowling approximation is dropped, obtained with a nonlinear code by Zink et al. \cite{Zink10}. They use polytropic EoS and their model S has a polytropic index of $\Gamma=2$ while for their model C it is $\Gamma=2.5$. As we can see the deviations from the Cowling data for the stable branches can be large for high rotation rates, but the data for the potentially unstable branches fit very well with our relations. This is a strong justification for the use of the Cowling approximation.

The differences between polytropes and realistic equations of state are more pronounced when we look at the relations for mode frequencies in the nonrotating limit. In Figure \ref{Fig:StaticFreq} we plotted the linear fit obtained for polytropes in the Cowling approximation as provided in \cite{Gaertig10} (dotted line), and the corresponding relation obtained in \cite{Andersson98a}, where both full GR and realistic EoS are considered (dash-dotted line) \footnote{Our relations \eqref{eq:sigmaZeroNonrot2} -- \eqref{eq:sigmaZeroNonrot4} (solid lines) always lead to larger frequencies when compared to full GR \cite{Andersson98a}, which is due to the fact that the Cowling approximation is overestimating the $f$-mode frequencies.}.
The fit for polytropes clearly shows a different slope compared to most of the realistic EoS; similar differences between realistic and polytropic EoS were also observed in the full GR case \cite{Andersson98a, Andersson98c}.

As a conclusion we can say that the relations \eqref{eq:sigmaRotUnstFit_l2}--\eqref{eq:sigmaRotStableFit_l234} presented here are quite robust and do not depend significantly on the actual equation of state used. This behaviour can be attributed to the fact that the relations for the mode frequencies are normalized by their corresponding value in the nonrotating case which seems to properly mask the EoS-specific influence up to a large extent. We also expect that these relations will approximately remain valid even if the Cowling approximation is dropped as it is indicated by the full GR results depicted in Figure \ref{Fig:sigmaComoving_l=234}.

\subsubsection{Asteroseismology relations for damping times}\label{asteroForDampingTimes}

As discussed in detail above, it is not possible to directly obtain gravitational wave damping times from simulations performed in the Cowling approximation. Instead equations \eqref{eq:dampingTime} -- \eqref{eq:Jlm} are employed which are based on the multipole formulae.

Evaluating these relations numerically turned out to be a bit intricate when using realistic EoS. First, as already mentioned above, realistic EoS typically exhibit sharp drops in the speed of sound close to the neutron drip density
which lead to numerical instabilities in the time integration of the perturbation equations. This can be attenuated by using a higher resolution of the computational domain compared to polytropes but still it is more difficult to obtain smooth eigenfunctions especially below the neutron drip density. Second, we are not directly evolving the fluid perturbation variables but some combinations of them -- the $Q_i$ variables, see eqs. \eqref{eq:Q_variables}. In order to reconstruct the primitive fluid perturbations one typically has to divide by $(\epsilon + p)$ at some point. Both the pressure and the energy density of neutron stars are negligibly small in the region of the outer crust compared to the corresponding values in the core, for all of the studied realistic EoS and this introduces large errors in the derived fluid perturbation variables. Thus the combination of these issues leads to large errors especially in the perturbations of the fluid four-velocity in neutron star regions below the neutron drip point. Since for compact objects with realistic EoS  the outer crust contains only a small portion of the stellar mass, due to the comparatively low density there, neglecting this region would not have a significant impact in evaluating the damping time relations \eqref{eq:dampingTime} -- \eqref{eq:Jlm}. We therefore neglect this region when computing the integrals and choose a cutoff density of $\sim10^{12} {\rm g/cm^3}$. The results show that the damping time does not change more than 10\% when this cutoff density is increased or decreased by approximately one order or magnitude.

As already mentioned, in the Cowling approximation mode frequencies can be  overestimated by up to 30 \%. As one can see from equations \eqref{eq:EnergyModes}--\eqref{eq:Jlm}, the energy loss due to gravitational radiation is proportional to $\omega^{2l+2}$ while the energy of the mode scales proportional to $\omega^2$. Therefore
the damping time\footnote{These relations are given in the case when $\Omega=0$, i.e. when $\omega_c=\omega_i=\omega$, for simplicity.} should be proportional to $\omega^{-2l}$. This means that rather small deviations in the frequencies can lead to large deviations in the corresponding damping times. Our results show that typically the damping times in the Cowling approximation are underestimated by up to a factor of 3.

Similar to the empirical relations for mode frequencies found in the previous Section \ref{asteroForFrequencies}, here we will derive two sets of relations for damping times which can be used for asteroseismology -- one set determines the functional behaviour of the normalized damping times with increasing rotation rate and the second set relates the normalization factor, i.e. the damping time in the nonrotating limit, to mass and radius of the star.

\begin{figure}[ht!]
\centering
\includegraphics[width=0.55\textwidth]{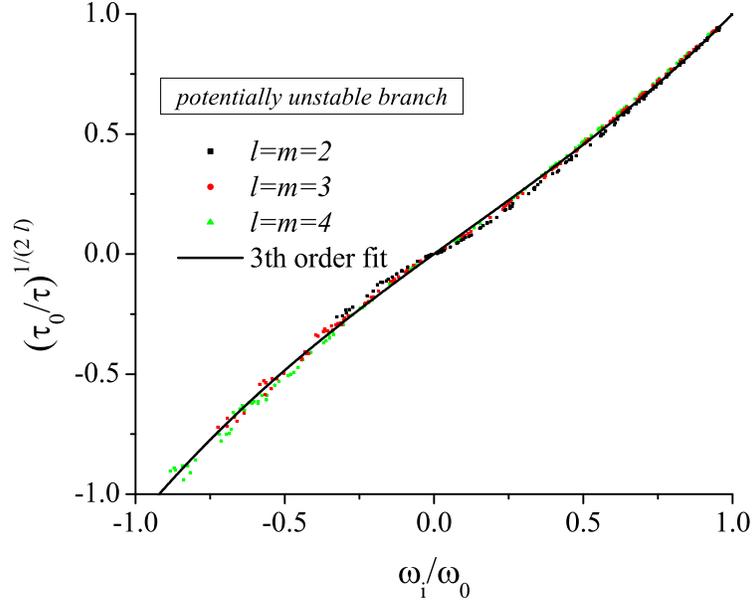}
\caption{Normalized damping times $(\tau_0/\tau)^{1/2l}$ as a function of normalized mode frequencies  in the inertial frame $\omega_i/\omega_0$ for potentially unstable branches and  $l=m=2, 3, 4$.}
\label{Fig:DampingUnstable}
\end{figure}

In Figure \ref{Fig:DampingUnstable}, normalized damping times of the potentially unstable branches as a function of normalized mode frequencies in the inertial frame are depicted for $l = 2, 3, 4$ and for all EoS used in this study. The quantities are normalized to their corresponding values in the nonrotating limit. On the ordinate we plot $(\tau_0/\tau)^{1/2l}$ because $\tau \sim \omega^{-2l}$ as discussed above. It is convenient to use inertial frame mode frequencies on the abscissa because there is a one-to-one mapping between rotation rates and mode frequencies in this case.

In order to fit the data, a third order polynomial is used similar to the approach in \cite{Gaertig10}. If we assume that this polynomial is of the form $y(x) = A + B x + C x^2 + D x^3$ and since normalized quantities are used, we require that $y(x)|_{x=1}=1$. Also since $\tau \sim \omega^{-2l}$, we can conclude that the combination $\tau_0/\tau$ vanishes when a neutral mode appears in the inertial frame. This means that a second constraint can be imposed on the fitting polynomial, namely $y(x)|_{x=0}=0$. In order to fulfill these constraints one can therefore choose $A=0$ and $B = 1-C-D$ by which we are left with only two independently adjustable parameters of the fit, i.e. the coefficients $C$ and $D$. As one can see from Figure \ref{Fig:DampingUnstable}, the data for all values of $l$ considered here can be approximated very well with a single polynomial and we obtain the following relation for the damping times of the potentially CFS-unstable modes

\begin{equation}
\label{eq:tauRotUnStab2}
\frac{\tau_0}{\tau} = {\rm sgn}(\omega_i^u) \left( 0.900 \left(\frac{\omega_i^u}{\omega_0}\right) - 0.057
 \left(\frac{\omega_i^u}{\omega_0}\right)^2 + 0.157 \left(\frac{\omega_i^u}{\omega_0}\right)^3\right)^{2l}\;,
\end{equation}
where ${\rm sgn}$ is the sign function.

As we pointed out, for the unstable branch it is $1/\tau \rightarrow 0$ when $\omega_i \rightarrow 0$. This constraint determines one of the free coefficients in the polynomial fit and facilitates to approximate the damping times for all values of the spherical index $l$ with a single fit. This approach is no longer applicable for the stable branch where both $\omega_i$ and $\omega_c$ are always nonzero and only the normalization condition $y(x)|_{x=1} = 1$ can be imposed. For the fitting of the stable branch we will use mode frequencies in the comoving frame which are a monotonic function of the stellar rotation rate in contrast to the corresponding frequencies in the inertial frame. We also plot $\tau/\tau_0$ for the stable branch on the ordinate because $\tau$ decreases while increasing the rotation rate in this case.

\begin{figure}[ht!]
\centering
\includegraphics[width=0.55\textwidth]{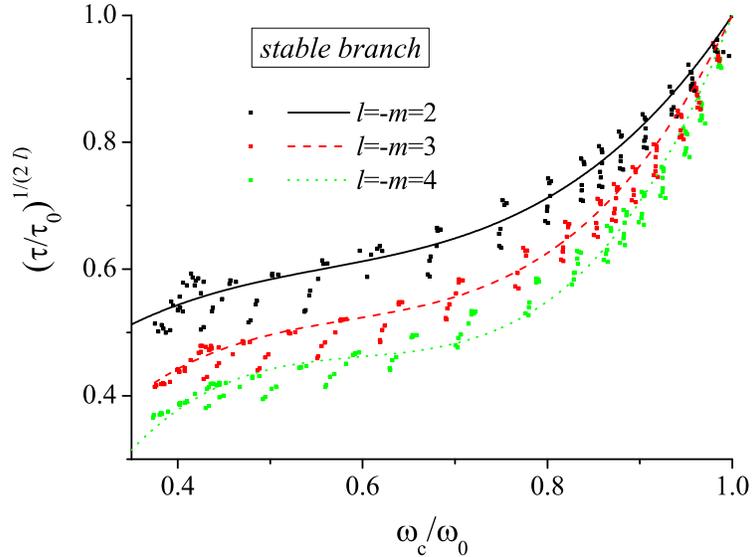}
\caption{Normalized damping times $(\tau/\tau_0)^{1/2l}$ as a function of normalized mode frequencies in the comoving frame $\omega_c/\omega_0$ for stable branches and $l= -m = 2, 3, 4$.}
\label{Fig:DampingStable}
\end{figure}

In Figure \ref{Fig:DampingStable}, normalized damping times of the stable branches as function of normalized mode frequencies in the comoving frame are depicted for $l = 2, 3, 4$ and for all EoS used in this study. Instead of the combination $\tau/\tau_0 \sim (\omega_c/\omega_0)$ which was used in \cite{Gaertig10}, here we use $\tau/\tau_0 \sim (\omega_c/\omega_0)^{2l}$; this turns out to be a more robust choice for the case $l > 2$.

This time, the normalization constraint  $y(x)|_{x=1} = 1$ leads to the relation $A = 1- B - C - D$, leaving three adjustable parameters of the fit. The results show that the spread in the data points for different EoS and different values of $l$ is larger than for the potentially unstable branch. Therefore, we choose to provide separate empirical relations for every spherical index $l = 2,3,4$.

After performing the polynomial fit, we obtain

\noindent
for $l = 2$:
\begin{equation}
\label{eq:tauRotStab2}
\left(\frac{\tau_{\;l=2}}{\tau_0}\right)^{1/4} =  -0.127 + 3.264 \left(\frac{\omega_c}{\omega_0}\right) -5.486 \left(\frac{\omega_c}{\omega_0}\right)^2 + 3.349 \left(\frac{\omega_c}{\omega_0}\right)^3\;,
\end{equation}
for $l = 3$:
\begin{equation}
\label{eq:tauRotStab3}
\left(\frac{\tau_{\;l=3}}{\tau_0}\right)^{1/6} = -0.672 + 5.270 \left(\frac{\omega_c}{\omega_0}\right) - 9.234 \left(\frac{\omega_c}{\omega_0}\right)^2 + 5.635 \left(\frac{\omega_c}{\omega_0}\right)^3\;,
\end{equation}
for $l = 4$:
\begin{equation}
\label{eq:tauRotStab4}
\left(\frac{\tau_{\;l=4}}{\tau_0}\right)^{1/8} =  -1.227 + 7.520 \left(\frac{\omega_c}{\omega_0}\right) - 13.500 \left(\frac{\omega_c}{\omega_0}\right)^2 + 8.207 \left(\frac{\omega_c}{\omega_0}\right)^3\;.
\end{equation}

Since in all the relations for damping times above only normalized quantities are used, there is the reasonable expectation that although the Cowling approximation is used here, the functional form of the empirical relations will remain valid even if this approximation is dropped.

\begin{figure}[ht!]
\centering
\includegraphics[width=0.45\textwidth]{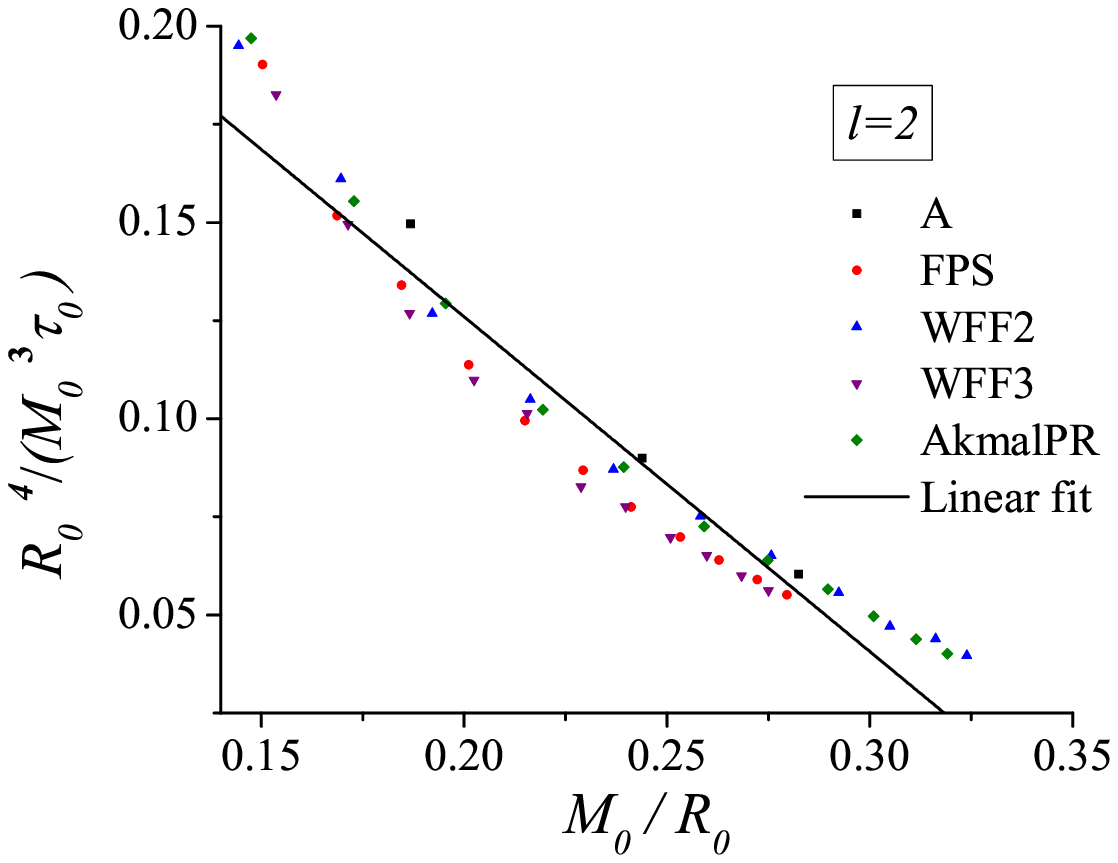}
\includegraphics[width=0.49\textwidth]{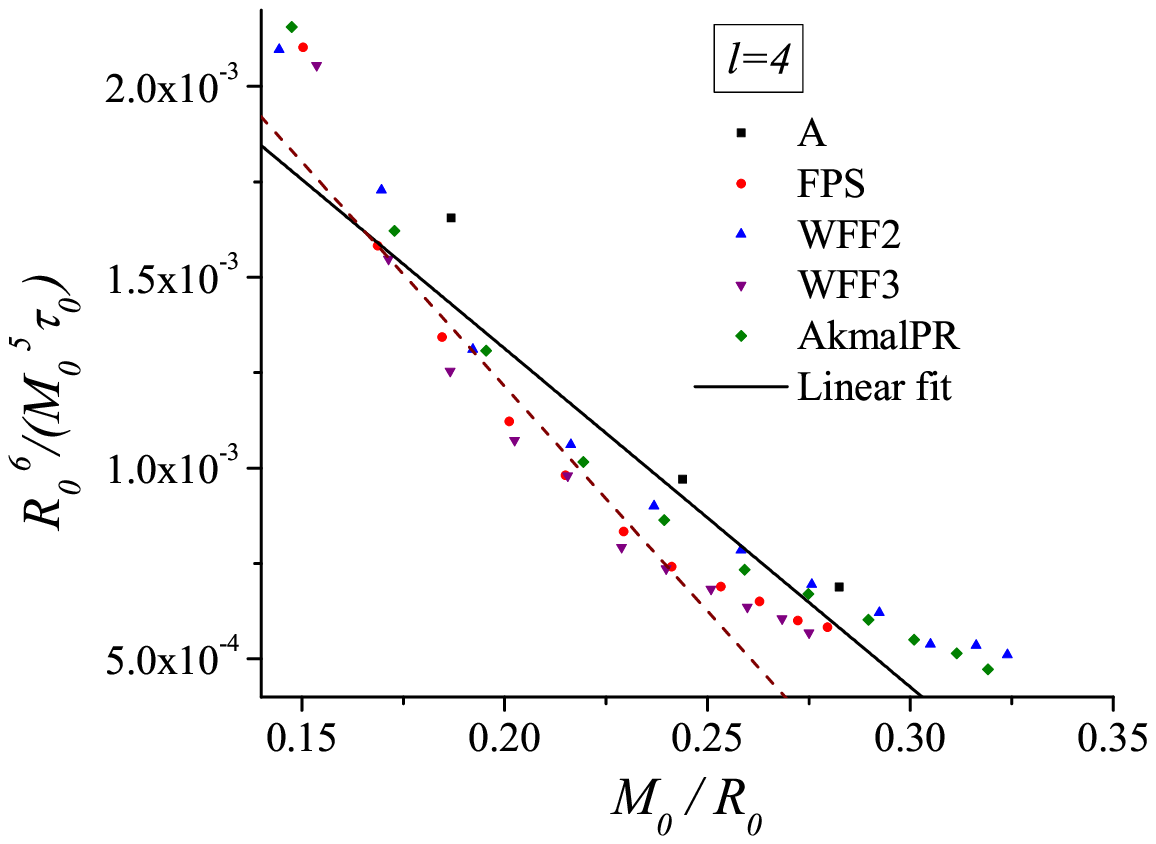}
\caption{Normalized damping times as function of the compactness $M/R$ of the star. The results for $l=2, 4$ are depicted for all realistic EoS used in this study. }
\label{Fig:StaticDamping_l=2_4}
\end{figure}

Finally, the relations for damping times in the static limit as function of mass and radius are needed. One can show that a rough estimate for the damping time, given by the quadrupole formula, is  $\tau_0 \sim \bar{R} (\bar{R}/\bar{M})^{l+1}$  \cite{Detweiler75,Andersson98a} and this relation can be used for the normalization of $\tau_0$. Thus in Figure \ref{Fig:StaticDamping_l=2_4}, we plot damping times as a function of the compactness $M/R$ of the star, where the damping time is normalized by $\bar{R} (\bar{R}/\bar{M})^{l+1}$.

Performing a linear fit of the static neutron star damping times, we obtain

\noindent
for $l=2$:
\begin{equation}
\label{eq:tauStatic2}
\frac{1}{\tau_0\,\mathrm{[s]}} = \frac{\bar{M}_0^3}{\bar{R}_0^4}\left[78.55  - 46.71 \left(\frac{{\bar M}_0}{{\bar R}_0}\right)\right]\;,
\end{equation}
for $l=3$:
\begin{equation}
\label{eq:tauStatic3}
\frac{1}{\tau_0\,\mathrm{[s]}} = \frac{\bar{M}_0^4}{\bar{R}_0^5}\left[1.691  - 1.027 \left(\frac{{\bar M}_0}{{\bar R}_0}\right)\right]\;,
\end{equation}
and for $l=4$:
\begin{equation}
\label{eq:tauStatic4}
\frac{1}{\tau_0\,\mathrm{[s]}} = \frac{\bar{M}^5_0}{\bar{R}^6_0}\left[0.0350  - 0.0208 \left(\frac{{\bar M}_0}{{\bar R}_0}\right)\right]\;.
\end{equation}
Similar to the corresponding relations for mode frequencies, these fittings are most sensitive to the deviations introduced by the Cowling approximation.

\begin{figure}[ht!]
\centering
\includegraphics[width=0.49\textwidth]{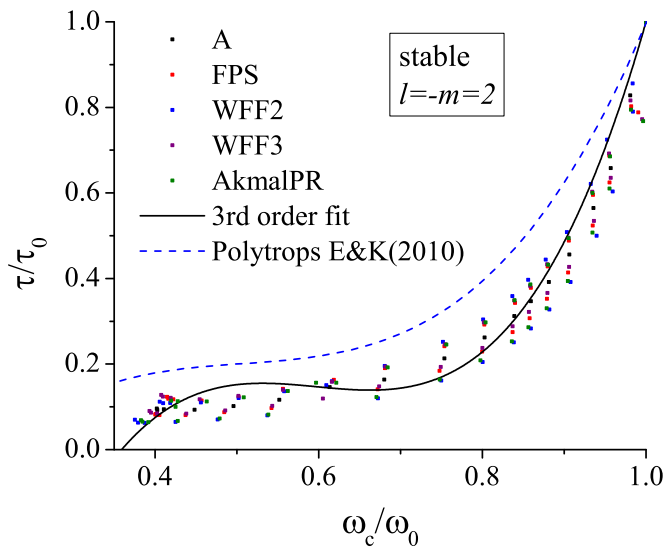}
\caption{Normalized damping times $\tau/\tau_0$ as function of normalized mode frequencies in the comoving frame $\omega_c/\omega_0$. The analytic dependence for polytropes found in \cite{Gaertig10} is depicted as dashed line. }
\label{Fig:DampingHigher_ComparePoly}
\end{figure}

Let us again draw a comparison with polytropic EoS at this point. The relation for the normalized damping times of the potentially unstable branch \eqref{eq:tauRotUnStab2} is quite similar in both cases, due to the fact that there are only two independently adjustable parameters. The corresponding relation for the stable branches changes though. First, it was already pointed out that the relations used here \eqref{eq:tauRotStab2}--\eqref{eq:tauRotStab4} differ slightly from the ones in \cite{Gaertig10} -- here we plot $\tau/\tau_0 \sim (\omega_c/\omega_0)^{2l}$ instead of  $\tau/\tau_0 \sim (\omega_c/\omega_0)$. In order to compare our results for realistic EoS with the polytropic ones, the dependence  $\tau/\tau_0 \sim (\omega_c/\omega_0)$ for $l=-m=2$ is depicted in Figure \ref{Fig:DampingHigher_ComparePoly}. The analytic dependence for polytropes  found in \cite{Gaertig10} is shown there as well. As one can see, the difference is quite big, but this is most likely due to the fact that in \cite{Gaertig10} several very soft equations of state are used while most of the realistic EoS utilized here are rather stiff. If one would exclude the very soft EoS from \cite{Gaertig10}, the relations for both the polytropes and the realistic EoS will be quite similar.

When comparing the damping times between polytropes and realistic EoS in the nonrotating limit, we have to keep in mind the following: Due to the errors in the damping times related to the Cowling approximation, a correction factor was introduced in \cite{Gaertig10} in order to compensate the deviations from full GR. This factor was derived after a systematic comparison with fully relativistic results in the quadrupolar case \cite{Balbinski85}. Since we also present relations for $l>2$ here, the correction factor is unknown so we decided to present the original results for the damping time.

\begin{figure}[ht!]
\centering
\includegraphics[width=0.5\textwidth]{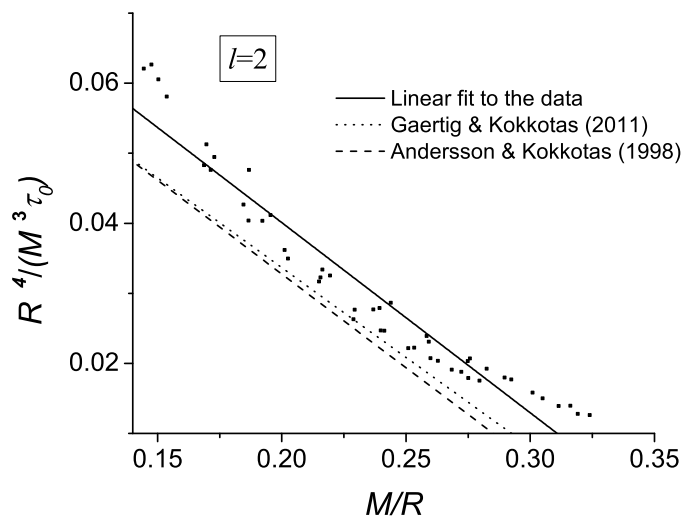}
\caption{Normalized damping times as function of the compactness $M/R$ for nonrotating models. The correction factor used in \cite{Gaertig10} is introduced in order to compare our data to the polytropic case presented there.}
\label{Fig:DampingStatic_ComparePoly}
\end{figure}

In Figure \ref{Fig:DampingStatic_ComparePoly},  we show the fits for damping times of quadrupolar modes in the nonrotating limit for both polytropes and realistic EoS. In order to make a proper comparison, we introduce the same correction factor used in \cite{Gaertig10}. As one can see, the fit for realistic EoS generally leads to smaller damping times compared to polytropes. This might be due to the fact that the correction factor for our set of EoS is different from the one used for polytropes.\footnote{Strictly speaking this factor does not only depend on the EoS, but most likely on mass and radius of the stars as well.}. Another possible source of error might be our treatment of the numerical instabilities near the neutron drip point, see the discussion at the beginning of this Section \ref{asteroForDampingTimes}.

\section{Solving the inverse problem}\label{sec:InverseProblem}
After obtaining empirical relations for gravitational wave asteroseismology, we need to address the inverse problem -- determining the mass, radius and rotation rate of a neutron star when some observed frequencies and/or damping times are provided. Since three characteristic neutron star parameters need to be identified, one correspondingly needs three observables. But not any combination of frequencies and damping times is suitable for solving the inverse problem. For example, in the simplest case one could suggest to use three frequencies of different modes in order to determine neutron star parameters. But the empirical relations found for frequencies of fast rotating neutron stars can only be used to obtain the rotation rate $\Omega$ and the average density $M/R^3$ but not mass and radius independently. The reason is that in relations \eqref{eq:sigmaRotUnstFit_l2}--\eqref{eq:OmegaK} the independent variables are $\Omega$, $\Omega_K$ and $M/R^3$. Since $\Omega_K$ can also be expressed as a function of $M/R^3$ up to leading order, see \eqref{eq:OmegaK},\footnote{As mentioned above, the coefficients in the relation (\ref{eq:OmegaK}) also depend on the compactness $M/R$ but this is a second-order effect and cannot be used to accurately determine $M$ and $R$.} this cannot be used to provide an additional constraint on mass and radius. We are led to the conclusion that by observing at least two mode frequencies of a single rotating star we will be able to determine its rotation rate and average density. Of course detecting even more frequencies will aid to set additional constraints on these parameters and to provide robust error estimates. For example, solving the inverse problem sometimes can lead to more than just one physically feasible solution. Observing additional frequencies thus can facilitate to determine a unique solution for $\Omega$ and $M/R^3$.

In order to compute masses and radii independently and not just a mere combination of them, one needs to observe the damping time of at least one of the $f$-modes where the relevant empirical relations are given by \eqref{eq:tauRotUnStab2} -- \eqref{eq:tauStatic4}. Of course, observing the damping times of neutron star oscillations is supposed to be even more difficult than detecting the oscillation frequencies since the mode needs to be tracked for a substantial amount of time in the noisy detector data. An alternative way to determine mass and radius is to detect frequencies of other modes like $w$- or the $p$-modes, similar to the study in the nonrotating case \cite{Andersson98a}. But these oscillations are supposed to reach lower amplitudes, their frequency band lies outside the maximum sensitivity range of current detectors and they are damped away faster. The $r$-modes on the other hand are generically CFS unstable \cite{Andersson98b,Friedman98} and might be observed easier. On the other hand, $r$-modes form a dense spectrum, distributing energy very efficiently amongst them and most likely to other $p$-modes as well and therefore quickly drop in amplitude once they are excited. Their frequencies are up to leading order proportional to the rotation rate of the star \cite{Kokkotas99b} so they cannot be used to determine its mass and radius. However they might help to constrain the exact value of $\Omega$ even further. Since the main goal of this paper is to thoroughly study gravitational wave asteroseismology with $f$-modes, we will stick to these modes only.

We now proceed to give some typical asteroseismology examples which are divided into two classes. First, only two oscillation frequencies are used to obtain rotation rate and average density of the star. As discussed above, mode frequencies should be easier to detect with appropriate accuracy. In the second example, we use two oscillation frequencies and a single damping time in order to determine mass and radius  independently. Our results for solving the inverse problem are given in Table \ref{tbl:Seism_WFF2} and Table \ref{tbl:Seism_FPS}. In the results presented here we also apply the formulas \eqref{eq:Mnorm_RotCorrection} and \eqref{eq:Rnorm_RotCorrection} for the rotational correction of the mass and the radius.

\begin{table*}[ht!]
\begin{center}
\caption{Solutions of the inverse problem  using two frequencies for EoS WFF2. Models with two different masses in the nonrotating limit $M_0=1.4M_\odot$ ($\Omega_K/2\pi = 1.273$ kHz) and $M_0=2.0M_\odot$ ($\Omega_K/2\pi = 1.687$ kHz) are given, and for each mass we provide two rotation rates -- one rotating moderately fast and the other rotating close to the Kepler limit. The frequencies are measured in kHz and the percent deviations from the exact values are given in brackets}\label{tbl:Seism_WFF2}
\begin{tabular}{ccccccccc}
\hline
\hline
{\bf EoS:} & WFF2 &  & WFF2  &  & WFF2 &  & WFF2 & \\
{\bf Mass (nonrot.):} & 1.4 $M_\odot$ &  & 1.4 $M_\odot$ &  & 2.0 $M_\odot$ &  & 2.0 $M_\odot$ & \\
\hline
\hline
 &  &   &  &  &  &  &  & \\
 & ${\bar M}/{\bar R}^3$ & $\Omega/2\pi$ & ${\bar M}/{\bar R}^3$ & $\Omega/2\pi$ & ${\bar M}/{\bar R}^3$ & $\Omega/2\pi$ & ${\bar M}/{\bar R}^3$ & $\Omega/2\pi$\\
\hline
{\bf Exact} & 0.711 & 0.308 & 0.463 & 1.209 & 1.149 & 0.410 & 0.805 & 1.608\\[0.1in]
$\bm\omega_{\bm l\bm =\bm 3}^{\bm u}$\;\bf\&\; $\bm\omega_{\bm l \bm =\bm 2}^{\bm s}$ & 0.653 (8) & 0.305 (1) & 0.470 (2) & 1.187 (2) & 0.956 (17) & 0.418 (2) & 0.605 (25) & 1.635 (2)\\[0.1in]
$\bm\omega_{\bm l\bm =\bm 4}^{\bm u}$\;\bf\&\; $\bm\omega_{\bm l\bm =\bm 2}^{\bm s}$ & 0.650 (9) & 0.307 (0.3) & 0.468 (1) & 1.189 (2) & 0.963 (16) & 0.413 (1) & 0.607 (25) & 1.620 (1)\\[0.1in]
$\bm\omega_{\bm l \bm =\bm 3}^{\bm u}$ \;\bf \&\; $\bm\omega_{\bm l\bm =\bm 4}^{\bm u}$ & 0.720 (1) & 0.329 (7) & 0.565 (22) & 1.230 (2) & 0.780 (32) & 0.363 (11) & 0.267 (67) & 1.286 (20)\\[0.05in]
\hline
\hline
\end{tabular}
\end{center}
\end{table*}

\begin{table*}[ht!]
\begin{center}
\caption{Solutions of the inverse problem  using two frequencies for EoS FPS. Models with two different masses in the nonrotating limit $M_0=1.4M_\odot$ ($\Omega_K/2\pi = 1.315$ kHz) and $M_0=1.7M_\odot$ ($\Omega_K/2\pi = 1.628$ kHz) are given, and for each mass we provide two rotation rates -- one rotating moderately fast and the other rotating close to the Kepler limit. The frequencies are measured in kHz and the percent deviations from the exact values are given in brackets.}\label{tbl:Seism_FPS}
\begin{tabular}{cccccccccccc}
\hline
\hline
\bf EoS: & FPS &  & FPS  &  & FPS &  & FPS & \\
\bf Mass (nonrot.): & 1.4 $M_\odot$ &  & 1.4 $M_\odot$ &  & 1.7 $M_\odot$ &  & 1.7 $M_\odot$ & \\
\hline
\hline
 &  &   &  &  &  &  &  & \\
 & ${\bar M}/{\bar R}^3$ & $\Omega/2\pi$ & ${\bar M}/{\bar R}^3$ & $\Omega/2\pi$ & ${\bar M}/{\bar R}^3$ & $\Omega/2\pi$ & ${\bar M}/{\bar R}^3$ & $\Omega/2\pi$\\
\hline
\bf Exact & 0.764 & 0.397 & 0.546 & 1.195 & 1.115 & 0.494 & 0.749 & 1.565\\[0.1in]
$\bm\omega_{\bm l\bm = \bm 3}^{\bm u}$\;\bf\&\; $\bm\omega_{\bm l\bm = \bm 2}^{\bm s}$ &0.765 (0.1) & 0.390 (2) & 0.630 (15) & 1.164 (3) & 1.119 (0.4) & 0.496 (0.4) & 0.556 (26) & 1.562 (0.2)\\[0.1in]
$\bm\omega_{\bm l\bm = \bm4}^{\bm u}$\;\bf\&\; $\bm\omega_{\bm l\bm = \bm2}^{\bm s}$ &0.761 (0.4) & 0.395 (1) & 0.629 (15) & 1.166 (2) & 1.112 (0.3) & 0.500 (1) & 0.560 (25) & 1.559 (0.4)\\[0.1in]
$\bm\omega_{\bm l\bm =\bm 3}^{\bm u}$\;\bf\&\; $\bm\omega_{\bm l\bm = \bm 4}^{\bm u}$ &0.953 (25) & 0.448 (13) & 0.677 (24) & 1.201 (1) & 1.311 (18) & 0.546 (11) & 0.542 (28) & 1.501 (4)\\[0.05in]
\hline
\hline
\end{tabular}
\end{center}
\end{table*}

As representative examples, we choose the two mode frequencies to belong either to $l=m=3, 4$ oscillations or  a combination of one of the $l=m=3$ or $l=m=4$ oscillations and a $l=-m=2$ oscillation. This choice is motivated by the fact that the $l=m=3, 4$ modes are supposed to develop the secular CFS instability much earlier than quadrupolar oscillations.
The superscripts $u$ and $s$ for the mode frequencies refer to potentially unstable ($m>0$) and stable ($m<0$) modes respectively. In these two Tables, a large range of masses, equations of state and rotation rates is covered and the percent deviations from the exact values are given in brackets. Still, for most of the models, the rotation rate and the compactness can be recovered with a good accuracy.
Only the deviations for some of the more massive models with very high rotational rates could be large. This is due not only to inaccuracy in the asteroseismology formulas \eqref{eq:sigmaRotUnstFit_l2}--\eqref{eq:OmegaK}, but also to the rotational corrections \eqref{eq:Mnorm_RotCorrection} and \eqref{eq:Rnorm_RotCorrection} which may overestimate the mass and the radius of the star a lot in some cases.

The second set of asteroseismology examples is given in Tables \ref{tbl:SeismRot_WFF2} and \ref{tbl:SeismRot_FPS}.

\begin{table*}[ht!]
\begin{center}
\caption{Solution of the full inverse problem using two frequencies and a single damping time for EoS WFF2. Models with two different masses in the nonrotating limit $M_0=1.4M_\odot$ ($\Omega_K/2\pi = 1.273$ kHz) and $M_0=2.0M_\odot$ ($\Omega_K/2\pi = 1.687$ kHz) are given similar to Table \ref{tbl:Seism_WFF2}. The percent deviations from the exact values are given in brackets.}\label{tbl:SeismRot_WFF2}
\begin{tabular}{cccccccc}
\hline
\hline
\bf EoS: & WFF2 &  &  &  & WFF2  &  &  \\
\bf Mass (nonrotating): & 1.4$M_\odot$ &  &  &  & 1.4$M_\odot$ &  &  \\
\hline
\hline
 &  &  &  &  &  &  &  \\
 & $M [M_\odot]$ & $R [{\rm km}]$ & $\Omega/2\pi [{\rm kHz}]$ &  & $M [M_\odot]$ & $R [{\rm km}]$ & $\Omega/2\pi[{\rm kHz}]$ \\
\hline
\bf Exact & 1.41 & 11.23 & 0.308 &  & 1.72 & 13.84 & 1.209\\[0.1in]
$\bm\omega_{\bm l\bm = \bm 2}^{\bm s}$\;\bf\&\; $\bm\omega_{\bm l\bm = \bm 3}^{\bm u}$\;\bf\&\; $\bm\tau_{\bm l\bm = \bm 3}^{\bm u}$ & 1.54 (9) & 11.90 (6) & 0.305 (1) &  & 1.58 (8) & 13.37 (3) & 1.187 (2) \\[0.1in]
$\bm\omega_{\bm l\bm = \bm 2}^{\bm s}$\;\bf\&\; $\bm\omega_{\bm l\bm = \bm 4}^{\bm u}$\;\bf\&\; $\bm\tau_{\bm l\bm = \bm 4}^{\bm u}$ & 1.59 (13) & 12.05 (7) & 0.307 (0.3) &  & 2.51 (46) & 15.64 (13) & 1.189 (2) \\[0.1in]
$\bm\omega_{\bm l\bm = \bm 3}^{\bm u}$\;\bf\&\; $\bm\omega_{\bm l\bm = \bm 4}^{\bm u}$\;\bf\&\; $\bm\tau_{\bm l\bm = \bm 4}^{\bm u}$ & 1.63 (16) & 11.73 (4) & 0.329 (7) &  & 2.78 (62) & 15.21 (10) & 1.230 (2)\\[0.1in]
\hline
\hline
 &  &  &  &  &  &  &  \\
\bf EoS: & WFF2 &  &  &  & WFF2  &  &  \\
\bf Mass (nonrotating): & 2.0$M_\odot$ &  &  &  & 2.0$M_\odot$ &  &  \\
\hline
\hline
 &  &  &  &  &  &  &  \\
 & $M [M_\odot]$ & $R [{\rm km}]$ & $\Omega/2\pi [{\rm kHz}]$ &  & $M [M_\odot]$ & $R [{\rm km}]$ & $\Omega/2\pi[{\rm kHz}]$ \\
\hline
\bf Exact  & 2.02 & 10.79 & 0.410 &  & 2.38 & 12.83 & 1.608\\[0.1in]
$\bm\omega_{\bm l\bm = \bm 2}^{\bm s}$\;\bf\&\; $\bm\omega_{\bm l\bm = \bm 3}^{\bm u}$\;\bf\&\; $\bm\tau_{\bm l\bm = \bm 3}^{\bm u}$ & 1.98 (2) & 11.20 (4) & 0.418 (2) &  & 2.41 (1) & 14.34 (12) & 1.635 (2) \\[0.1in]
$\bm\omega_{\bm l\bm = \bm 2}^{\bm s}$\;\bf\&\; $\bm\omega_{\bm l\bm = \bm 4}^{\bm u}$\;\bf\&\; $\bm\tau_{\bm l\bm = \bm 4}^{\bm u}$ & 2.24 (11) & 11.71 (9) & 0.413 (1) &  & 2.70 (13) & 14.87 (16) & 1.620 (1) \\[0.1in]
$\bm\omega_{\bm l\bm = \bm 3}^{\bm u}$\;\bf\&\; $\bm\omega_{\bm l\bm = \bm4}^{\bm u}$\;\bf\&\; $\bm\tau_{\bm l\bm = \bm 4}^{\bm u}$ & 2.04 (1) & 12.33 (14) & 0.363 (11) &  & 2.23 (6) & 18.16 (42) & 1.286 (20) \\[0.05in]
\hline
\hline
\end{tabular}
\end{center}
\end{table*}

\begin{table*}[ht!]
\begin{center}
\caption{Solutions of the full inverse problem using two frequencies and a single damping time for EoS FPS. Models with two different masses in the nonrotating limit $M_0=1.4M_\odot$ ($\Omega_K/2\pi = 1.315$ kHz) and $M_0=1.7M_\odot$ ($\Omega_K/2\pi = 1.628$ kHz) are given similar to Table \ref{tbl:Seism_FPS}. The percent deviations from the exact values are given in brackets.}\label{tbl:SeismRot_FPS}
\begin{tabular}{cccccccc}
\hline
\hline
\bf EoS: & FPS &  &  &  & FPS  &  &  \\
\bf Mass (nonrotating): & 1.4$M_\odot$ &  &  &  & 1.4$M_\odot$ &  &  \\
\hline
\hline
 &  &  &  &  &  &  &  \\
 & $M [M_\odot]$ & $R [{\rm km}]$ & $\Omega/2\pi [{\rm kHz}]$ &  & $M [M_\odot]$ & $R [{\rm km}]$ & $\Omega/2\pi[{\rm kHz}]$ \\
\hline
\bf Exact & 1.42 & 10.99 & 0.397 &  & 1.64 & 12.90 & 1.195\\[0.1in]
$\bm\omega_{\bm l\bm = \bm 2}^{\bm s}$\;\bf\&\; $\bm\omega_{\bm l\bm = \bm 3}^{\bm u}$\;\bf\&\; $\bm\tau_{\bm l\bm = \bm 3}^{\bm u}$ & 1.37 (4) & 10.86 (1) & 0.390 (2) &  & 0.99 (40) & 10.65 (17) & 1.164 (3)\\[0.1in]
$\bm\omega_{\bm l\bm = \bm 2}^{\bm s}$\;\bf\&\; $\bm\omega_{\bm l\bm = \bm 4}^{\bm u}$\;\bf\&\; $\bm\tau_{\bm l\bm = \bm 4}^{\bm u}$ & 1.46 (3) & 11.12 (1) & 0.395 (1) &  & 1.88 (15) & 13.18 (2) & 1.166 (2)\\[0.1in]
$\bm\omega_{\bm l\bm = \bm 3}^{\bm u}$\;\bf\&\; $\bm\omega_{\bm l\bm = \bm4}^{\bm u}$\;\bf\&\; $\bm\tau_{\bm l\bm = \bm 4}^{\bm u}$ & 1.55 (9) & 10.51 (4) & 0.448 (13) &  & 1.92 (17) & 12.65 (2) & 1.201 (1)\\[0.1in]
\hline
\hline
 &  &  &  &  &  &  &  \\
\bf EoS: & FPS &  &  &  & FPS  &  &  \\
\bf Mass (nonrotating): & 1.7$M_\odot$ &  &  &  & 1.7$M_\odot$ &  &  \\
\hline
\hline
 &  &  &  &  &  &  &  \\
 & $M [M_\odot]$ & $R [{\rm km}]$ & $\Omega/2\pi [{\rm kHz}]$ &  & $M [M_\odot]$ & $R [{\rm km}]$ & $\Omega/2\pi[{\rm kHz}]$ \\
\hline
\bf Exact  & 1.72 & 10.33 & 0.494 &  & 2.01 & 12.42 & 1.565\\[0.1in]
$\bm\omega_{\bm l\bm = \bm 2}^{\bm s}$\;\bf\&\; $\bm\omega_{\bm l\bm = \bm 3}^{\bm u}$\;\bf\&\; $\bm\tau_{\bm l\bm = \bm 3}^{\bm u}$ & 2.39 (39) & 11.51 (11) & 0.496 (0.4) &  & 1.87 (7) & 12.46 (0.3) & 1.562 (0.2) \\[0.1in]
$\bm\omega_{\bm l\bm = \bm 2}^{\bm s}$\;\bf\&\; $\bm\omega_{\bm l\bm = \bm 4}^{\bm u}$\;\bf\&\; $\bm\tau_{\bm l\bm = \bm 4}^{\bm u}$ & 1.57 (9) & 10.04 (3) & 0.500 (1) &  & 2.40 (19) & 13.50 (9) & 1.559 (0.4)\\[0.1in]
$\bm\omega_{\bm l\bm = \bm 3}^{\bm u}$\;\bf\&\; $\bm\omega_{\bm l\bm = \bm4}^{\bm u}$\;\bf\&\; $\bm\tau_{\bm l\bm = \bm 4}^{\bm u}$ & 1.78 (3) & 9.90 (4) & 0.546 (11) &  & 2.11 (5) & 14.05 (13) & 1.501 (4)\\[0.05in]
\hline
\hline
\end{tabular}
\end{center}
\end{table*}

Here, the models from the previous example were used, but a single damping time of an unstable mode has been added as an additional input parameter.
In these tables the percent deviations from the exact mass, radius and rotational frequency are also shown.
As one can see, the error in finding mass and radius for some of the models and input data can be large, but still most of the examples provide quite good results.
Actually this accuracy can be improved even further by performing the scheme presented here in an iterative way. As we have seen, the first iteration already provided accurate estimates about masses, radii and rotation rates. With this information at hand, one can exclude certain EoS and repeat the fitting procedure by also narrowing down the allowed range of rotation rates that are consistent with the results from the first iteration. This will lead to more accurate empirical relations and also to a better convergence in the nonlinear root-finder, therefore leading to better estimates for the neutron star parameters. It is clear that this scheme can be repeated as often as necessary.

As an example we present the results obtained by performing a second iteration for the model with WFF2 EoS, $M=1.72 M_\odot$ and $\Omega/2\pi=1.209$ kHz. Our investigations show that a good strategy is to rederive only the relations for the normalized frequencies and damping times of rotating neutron stars (eqs. \eqref{eq:sigmaRotUnstFit_l2}--\eqref{eq:sigmaRotStableFit_l234}, \eqref{eq:tauRotUnStab2}) using data close to the computed value of $\Omega/\Omega_K$, and the relations for the nonrotating frequencies and damping times should remain the same. The reason is that sometimes the error in the average density and the compactness obtained after the first iteration could be large and rederiving the fits around these values could eventually make the results even more unprecise. In the next iterations though the static relations could be also refined because as we can see below, the error in determining $M$ and $R$ could be reduced significantly after the first iteration.

The results for the mass, the radius and the rotational frequency obtained after the second iteration are shown in Table \ref{tbl:SeismRot_SecondIter}. For most of the cases the second iteration leads to smaller errors. The  large deviations in $M$ and $R$, observed in Table \ref{tbl:SeismRot_WFF2} for some of the input data, were also reduced significantly. But we have to note that for very few cases it could happen that the second iteration does not improve the results and it can even increase the deviation a little bit. This happens especially for the high mass models with frequencies close to the Kepler limit where the errors in the fitting formulas are generally larger.

\begin{table*}[ht!]
\begin{center}
\caption{Results from the second iteration of solving the inverse problem for EoS WFF2, $M=1.4 M_\odot$ in the nonrotating limit and $\Omega/2\pi=1.209$ ($\Omega_K/2\pi = 1.273$ kHz). The results from the first iteration are presented in Table \ref{tbl:SeismRot_WFF2}. The percent deviations from the exact values are given in brackets.}\label{tbl:SeismRot_SecondIter}
\begin{tabular}{cccc}
\hline
\hline
\bf EoS: &  WFF2  &  &  \\
\bf Mass (nonrotating): & 1.4$M_\odot$ &  &  \\
\hline
\hline
 &  &  &  \\
 & $M [M_\odot]$ & $R [{\rm km}]$ & $\Omega/2\pi [{\rm kHz}]$ \\
\hline
\bf Exact & 1.72 & 13.84 & 1.209\\[0.1in]
$\bm\omega_{\bm l\bm = \bm 2}^{\bm s}$\;\bf\&\; $\bm\omega_{\bm l\bm = \bm 3}^{\bm u}$\;\bf\&\; $\bm\tau_{\bm l\bm = \bm 3}^{\bm u}$ &  1.89 (10) & 14.36 (4) & 1.202 (1) \\[0.1in]
$\bm\omega_{\bm l\bm = \bm 2}^{\bm s}$\;\bf\&\; $\bm\omega_{\bm l\bm = \bm 4}^{\bm u}$\;\bf\&\; $\bm\tau_{\bm l\bm = \bm 4}^{\bm u}$ &  1.91 (11) & 14.41 (4) & 1.204 (0.4) \\[0.1in]
$\bm\omega_{\bm l\bm = \bm 3}^{\bm u}$\;\bf\&\; $\bm\omega_{\bm l\bm = \bm 4}^{\bm u}$\;\bf\&\; $\bm\tau_{\bm l\bm = \bm 4}^{\bm u}$ &  1.98 (15) & 13.07 (6) & 1.270 (5) \\[0.05in]
\hline
\hline
\end{tabular}
\end{center}
\end{table*}

\section{Instability window}\label{sec:InstabWindow}
The last problem we are going to address here is the $f$-mode instability window for realistic equations of state (the case of polytropic equation of state was studied in \cite{Gaertig11,Passamonti12}). The instability window is the limiting curve in a  $T-\Omega$--representation where the exponential growth due to a CFS-unstable mode overcomes the dissipative effects, i.e. where the total damping time given by eq. \eqref{eq:fullTau} stays negative. As the neutron star evolves (cools down) it will generally move towards lower temperatures and rotational rates. When constructing the instability window one can no longer consider a sequence of models with fixed central energy density and increasing rotational rate -- we have to consider a sequence of models with constant baryon mass instead because this is the quantity that remains constant during the evolution of a single neutron star.

As explained above, we are working in the linear perturbation regime and we will take into account the following two viscous dissipation mechanisms --  bulk viscosity, which operates at high temperatures, and neutron shear viscosity, which damps out the oscillations at low temperatures. The relevant relations and coefficients are given in Section \ref{sec:DampingTimeFormulas}.

We computed the $l=m=2, 3, 4$ $f$-mode instability window for two sequences of constant baryon mass -- a sequence with the AkmalPR equation of state and a mass of $M=2.0\,M_\odot$ in the nonrotating limit, and a sequence with the WFF2 equation of state and the same mass $M=2.0\,M_\odot$ in the static case. We choose this particular configurations because both the AkmalPR and the WFF2 EoS support maximum masses above two solar masses, which is required from current neutron star observations \cite{Demorest10,Manousakis12,Antoniadis13}. Also, massive models are more compact as well and get CFS-unstable at lower rotation rates.

\begin{figure}[ht!]
\centering
\includegraphics[width=0.49\textwidth]{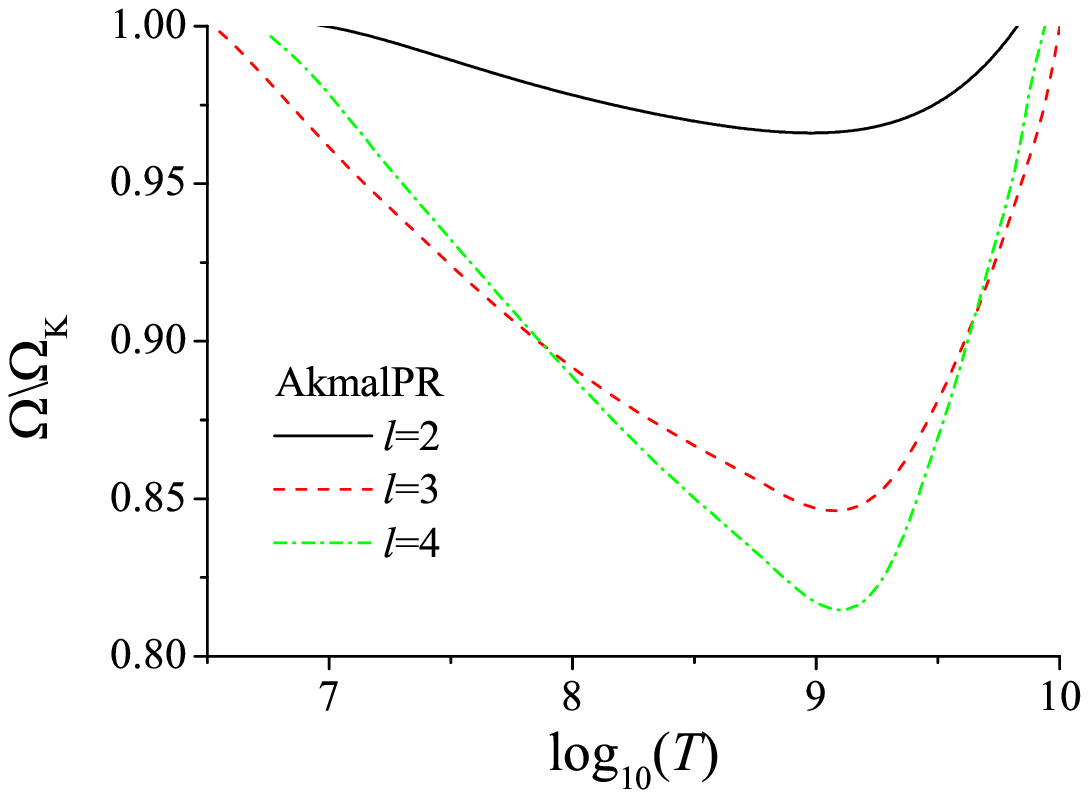}
\includegraphics[width=0.49\textwidth]{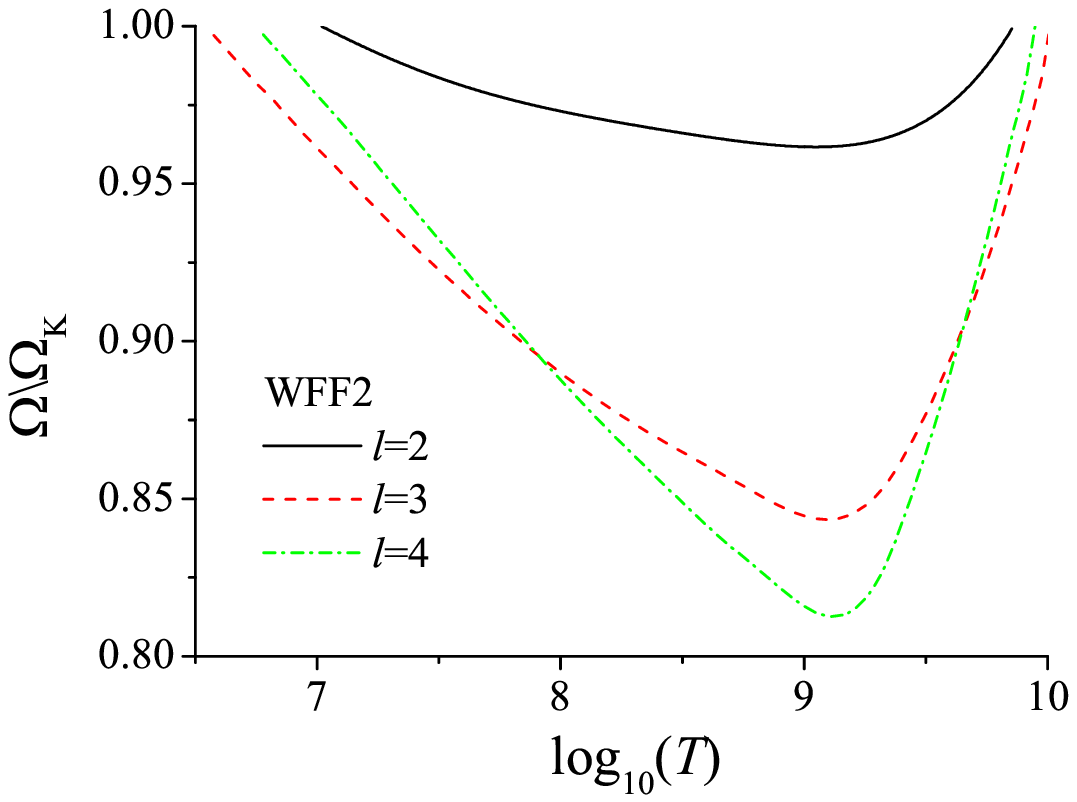}
\caption{Instability window for the AkmalPR and WFF2 EoS. The gravitational mass in the nonrotating limit is $M=2.0\,M_\odot$.}
\label{Fig:APR_WFF2InstabWindow}
\end{figure}

The $f$-mode instability window for the two EoS is depicted in Figure \ref{Fig:APR_WFF2InstabWindow} for $l=m=2, 3, 4$.\footnote{When constructing the instability window, we introduced a correction factor in the gravitational wave damping time similar to \cite{Gaertig10,Passamonti12}. This correction is required due to the Cowling approximation which underestimates damping times, but as it turns out the window does not change significantly even if the original results for the damping times are used.} It is evident that similar to the polytropic case, the quadrupolar modes are only marginally unstable -- the instability window reaches down to only about $96\%$ of the Kepler limit. More suitable candidates for detectable CFS-unstable modes are the cases with $l=3, 4$; there the instability window reaches down to $80-85\%$ of the Kepler limit. In this case a newborn and rapidly rotating neutron star may stay long enough in the instability windows during its evolution so that gravitational wave signals from the oscillations can be observed. It is important to note that for both equations of state the instability window is substantially deeper compared to all the polytropic models presented in \cite{Gaertig11,Passamonti12} which means that realistic equations of state might be more favourable to the CFS instability.

\begin{figure}[ht!]
\centering
\includegraphics[width=0.49\textwidth]{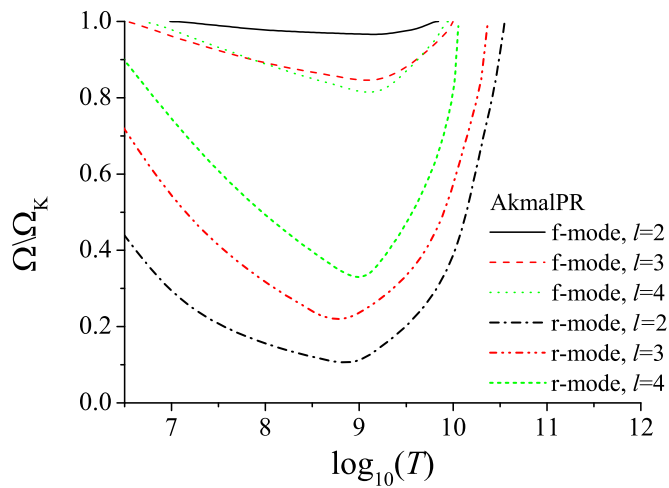}
\caption{Instability window for both $r$- and $f$-modes for the AkmalPR EoS. The gravitational mass in the nonrotating limit is $M = 2.0\,M_\odot$.} \label{Fig:InstabWindow_rmode}
\end{figure}

As it is well known, the $r$-modes are generically CFS-unstable \cite{Andersson98b,Friedman98}, i.e. they are unstable for any rotation rate of the star if additional dissipation mechanisms are neglected. Therefore the $r$-modes instability window will generally reach lower values of $\Omega/\Omega_K$, and it also covers a wider range of temperatures than the corresponding $f$-mode window. This can be seen in Figure \ref{Fig:InstabWindow_rmode}, where the instability window for both the $f$- and $r$-modes is plotted for the AkmalPR equation of state and $l=m=2, 3, 4$. One should keep in mind, that for computing the $r$-mode damping times, the current multipoles \eqref{eq:Jlm} are the dominant contribution to the energy loss. The frequencies and damping times computed with our time evolution code also match well with the analytic relations in  \cite{Kokkotas99b}.

As one can see, the instability window for the $r$-mode is much larger than for the $f$-mode for all of the considered values of the spherical index $l$ and as the neutron stars cool down, the $r$-mode will become unstable first. Thus the star will lose angular momentum quickly and may never reach the region of the $f$-mode instability. In practice though this scenario depends crucially on the $r$-mode  saturation amplitude -- if it is small enough then the star would lose angular momentum more slowly and it may eventually reach the $f$-mode instability window \cite{Passamonti12}. The results for nonlinear mode couplings of the $r$-modes suggest indeed that the saturation amplitude may be limited to small values \cite{Bondarescu07,Schenk02}. Thus if the saturation amplitude of the $f$-mode is large enough, the $f$-mode instability could develop in young neutron stars. Unfortunately, its saturation amplitude is still uncertain and further studies in this direction are needed to answer the question if CFS unstable $f$-modes of fast rotating neutron star can be observed.

\section{Conclusions}
In this paper we extended the results for nonaxisymmetric oscillations of fast rotating neutron stars in the Cowling approximation \cite{Gaertig08,Gaertig09,Gaertig10} by introducing realistic equations of state. We obtain the $f$-mode oscillation frequencies and damping times for a large set of equilibrium configurations with different EoS and central energy densities and then derive empirical relations that can be used for gravitational wave asteroseismology. We then study the inverse problem and at the end we consider the $f$-mode instability window for some models which are most promising to develop the CFS-instability. Another important aspect of our work is that the empirical relations obtained here are not only derived for the quadrupolar case but also for $l=|m|=3, 4$ as these modes get CFS unstable at lower rotation rates. This required some generalizations of the relations in \cite{Gaertig10} in order to be applicable for arbitrary values of $l$.

The results and the derived asteroseismology relations are compared with the polytropic ones presented in \cite{Gaertig10} and the following conclusions can be made. As we explained in detail in the previous section, the asteroseismology relations we derive can be divided into two groups -- relations for the normalized frequencies and damping times as a function of the rotational frequency and relations for the frequencies and damping times in the nonrotating limit. The first group of relations does not differ considerably from the polytropic case because we use normalized quantities. Moreover, by comparing with the few available full GR results \cite{Zink10} we show that these relations will be most probably very similar even if we drop the Cowling approximation. The biggest difference between polytropes and realistic EoS is in the second group of relations for the frequencies and the damping times in the nonrotating limit. More specifically the slope of the linear fit is different in the two cases which was also observed in the nonrotating full GR case \cite{Andersson98a, Andersson98c}.

Using the derived asteroseismology relations we extensively study the inverse problem -- what observational data is required in order to determine the mass, the radius and the rotational frequency of the star, and how accurately we can determine these parameters. We should note that our study is the first one to consider also the modes with $l>2$. It turns out that in order to solve the inverse problem we should be able to observe at least two frequencies and one damping time and the presented examples show that this information leads to good estimates for the neutron star parameters. For example we can observe the frequencies of $l=m=3$ and $l=m=4$ CFS unstable $f$-modes and one of the damping times of these modes, which is a realistic scenario.
The error in the parameter estimation can be reduced by performing at least one more iteration of solving the inverse problem, that is after having a first estimate of $M$, $R$ and $\Omega$ we can derive new asteroseismology relations valid for a smaller range of parameters when compared to the first estimate. Applying the refined relations once again to the input data generally results in a better accuracy for $M$, $R$ and $\Omega$.

If we are able to observe more than two modes this could also help us to set additional constraints on the mass and radius and also help us to determine the error bars. For example solving the inverse problem sometimes leads to more than one solution with physically reasonable values for the mass, radius and rotational frequencies and the additional observational information could help us to distinguish between these solutions.

We should keep in mind that all of the asteroseismology relations are derived within the Cowling approximation which introduced deviations in the $f$-modes oscillations frequencies and damping times. That is why it is important to drop this approximation and to consider the perturbations of the metric as well in the future. But the results in the present paper are valuable on their own because they show us the potential differences between the oscillations of neutron stars with polytropic and realistic EoS and how to consistently perform gravitational wave asteroseismology of fast rotating neutron stars for modes with higher values of $l>2$. Moreover, the comparison with the few available full GR data \cite{Zink10} suggests, that the normalized dependences for the fast rotating neutron stars will  remain similar even if we drop the Cowling approximation.

The last problem we address is the $f$-mode instability window for the AkmalPR and WFF2 equations of state and for rotational sequences with mass $M=2.0M_\odot$ in the nonrotating limit. Due to their high compactness, the chosen models are particularly good candidates to develop the CFS instability. As it is expected from the polytropic case \cite{Passamonti12,Gaertig11} the $l=m=3$ and $l=m=4$ modes can develop this instability for a much larger range of parameters compared to the $l=m=2$ modes. An important result is that the instability windows for these realistic EoS reaches lower bounds on the critical rotational frequency, where the CFS instability overcomes dissipative effects, than all the polytropic models presented in previous studies \cite{Gaertig11,Passamonti12}. But there is something else we have to take into account -- the $r$-mode instability window is in general much bigger, because the $r$-modes are CFS unstable for any rotational rate of the star. This is why we also calculate the $r$-mode instability window for one of the models in order to compare it to the corresponding $f$-mode window. The evolution of a newborn neutron star through the instability windows depends also heavily on the $r$ and $f$-mode saturation amplitudes \cite{Passamonti12} and further investigations in this direction are needed in order to answer the question if a $f$-mode CFS instability can develop in rapidly rotating neutron stars

\section*{Acknowledgements}
We would like to thank N. Stergioulas for providing his rns code and for the helpful discussions. We are grateful to S. Yazadjiev and B. Zink for the helpful discussions and advices. The realistic EoS were supplied by N. Stergioulas and by the LORENE code (http://www.lorene.obspm.fr/). DD acknowledges support from the German Science Foundation (DFG) via SFB/TR7 and by the Bulgarian National Science Fund under Grant
DMU-03/6. EG acknowledges support from the German Science Foundation (DFG) via SFB/TR7. CK acknowledges financial support from the EPSRC and the School of Mathematics of the University of Southampton.


\bibliography{references}

\begin{thebibliography}{65}
\expandafter\ifx\csname natexlab\endcsname\relax\def\natexlab#1{#1}\fi
\expandafter\ifx\csname bibnamefont\endcsname\relax
  \def\bibnamefont#1{#1}\fi
\expandafter\ifx\csname bibfnamefont\endcsname\relax
  \def\bibfnamefont#1{#1}\fi
\expandafter\ifx\csname citenamefont\endcsname\relax
  \def\citenamefont#1{#1}\fi
\expandafter\ifx\csname url\endcsname\relax
  \def\url#1{\texttt{#1}}\fi
\expandafter\ifx\csname urlprefix\endcsname\relax\def\urlprefix{URL }\fi
\providecommand{\bibinfo}[2]{#2}
\providecommand{\eprint}[2][]{\url{#2}}

\bibitem[{\citenamefont{Kokkotas and Schmidt}(1999)}]{Kokkotas99a}
\bibinfo{author}{\bibfnamefont{K.~D.} \bibnamefont{Kokkotas}} \bibnamefont{and}
  \bibinfo{author}{\bibfnamefont{B.~G.} \bibnamefont{Schmidt}},
  \bibinfo{journal}{\lrr} \textbf{\bibinfo{volume}{2}}, \bibinfo{pages}{2}
  (\bibinfo{year}{1999}).

\bibitem[{\citenamefont{Andersson}(2003)}]{Andersson03}
\bibinfo{author}{\bibfnamefont{N.}~\bibnamefont{Andersson}},
  \bibinfo{journal}{\cqg} \textbf{\bibinfo{volume}{20}}, \bibinfo{pages}{R105}
  (\bibinfo{year}{2003}).

\bibitem[{\citenamefont{Stergioulas}(2003)}]{Stergioulas03}
\bibinfo{author}{\bibfnamefont{N.}~\bibnamefont{Stergioulas}},
  \bibinfo{journal}{\lrr} \textbf{\bibinfo{volume}{6}}, \bibinfo{pages}{3}
  (\bibinfo{year}{2003}).

\bibitem[{\citenamefont{Andersson et~al.}(2011)\citenamefont{Andersson,
  Ferrari, Jones, Kokkotas, Krishnan et~al.}}]{Andersson10}
\bibinfo{author}{\bibfnamefont{N.}~\bibnamefont{Andersson}},
  \bibinfo{author}{\bibfnamefont{V.}~\bibnamefont{Ferrari}},
  \bibinfo{author}{\bibfnamefont{D.}~\bibnamefont{Jones}},
  \bibinfo{author}{\bibfnamefont{K.}~\bibnamefont{Kokkotas}},
  \bibinfo{author}{\bibfnamefont{B.}~\bibnamefont{Krishnan}},
  \bibnamefont{et~al.}, \bibinfo{journal}{\grg} \textbf{\bibinfo{volume}{43}},
  \bibinfo{pages}{409} (\bibinfo{year}{2011}).

\bibitem[{\citenamefont{Andersson and Kokkotas}(1998)}]{Andersson98a}
\bibinfo{author}{\bibfnamefont{N.}~\bibnamefont{Andersson}} \bibnamefont{and}
  \bibinfo{author}{\bibfnamefont{K.~D.} \bibnamefont{Kokkotas}},
  \bibinfo{journal}{\mnras} \textbf{\bibinfo{volume}{299}},
  \bibinfo{pages}{1059} (\bibinfo{year}{1998}).

\bibitem[{\citenamefont{Kokkotas et~al.}(2001)\citenamefont{Kokkotas,
  Apostolatos, and Andersson}}]{Kokkotas01}
\bibinfo{author}{\bibfnamefont{K.}~\bibnamefont{Kokkotas}},
  \bibinfo{author}{\bibfnamefont{T.}~\bibnamefont{Apostolatos}},
  \bibnamefont{and}
  \bibinfo{author}{\bibfnamefont{N.}~\bibnamefont{Andersson}},
  \bibinfo{journal}{\mnras} \textbf{\bibinfo{volume}{320}},
  \bibinfo{pages}{307} (\bibinfo{year}{2001}).

\bibitem[{\citenamefont{Benhar et~al.}(2004)\citenamefont{Benhar, Ferrari, and
  Gualtieri}}]{Benhar04}
\bibinfo{author}{\bibfnamefont{O.}~\bibnamefont{Benhar}},
  \bibinfo{author}{\bibfnamefont{V.}~\bibnamefont{Ferrari}}, \bibnamefont{and}
  \bibinfo{author}{\bibfnamefont{L.}~\bibnamefont{Gualtieri}},
  \bibinfo{journal}{\prd} \textbf{\bibinfo{volume}{70}},
  \bibinfo{pages}{124015} (\bibinfo{year}{2004}).

\bibitem[{\citenamefont{Gaertig and Kokkotas}(2011)}]{Gaertig10}
\bibinfo{author}{\bibfnamefont{E.}~\bibnamefont{Gaertig}} \bibnamefont{and}
  \bibinfo{author}{\bibfnamefont{K.~D.} \bibnamefont{Kokkotas}},
  \bibinfo{journal}{\prd} \textbf{\bibinfo{volume}{83}},
  \bibinfo{pages}{064031} (\bibinfo{year}{2011}).

\bibitem[{\citenamefont{Chandrasekhar}(1970)}]{Chandrasekhar70}
\bibinfo{author}{\bibfnamefont{S.}~\bibnamefont{Chandrasekhar}},
  \bibinfo{journal}{\prl} \textbf{\bibinfo{volume}{24}}, \bibinfo{pages}{611}
  (\bibinfo{year}{1970}).

\bibitem[{\citenamefont{Friedman and Schutz}(1978)}]{Friedman78}
\bibinfo{author}{\bibfnamefont{J.}~\bibnamefont{Friedman}} \bibnamefont{and}
  \bibinfo{author}{\bibfnamefont{B.~F.} \bibnamefont{Schutz}},
  \bibinfo{journal}{\apj} \textbf{\bibinfo{volume}{222}}, \bibinfo{pages}{281}
  (\bibinfo{year}{1978}).

\bibitem[{\citenamefont{Zink et~al.}(2010)\citenamefont{Zink, Korobkin,
  Schnetter, and Stergioulas}}]{Zink10}
\bibinfo{author}{\bibfnamefont{B.}~\bibnamefont{Zink}},
  \bibinfo{author}{\bibfnamefont{O.}~\bibnamefont{Korobkin}},
  \bibinfo{author}{\bibfnamefont{E.}~\bibnamefont{Schnetter}},
  \bibnamefont{and}
  \bibinfo{author}{\bibfnamefont{N.}~\bibnamefont{Stergioulas}},
  \bibinfo{journal}{\prd} \textbf{\bibinfo{volume}{81}},
  \bibinfo{pages}{084055} (\bibinfo{year}{2010}).

\bibitem[{\citenamefont{{Kojima}}(1993)}]{Kojima93}
\bibinfo{author}{\bibfnamefont{Y.}~\bibnamefont{{Kojima}}},
  \bibinfo{journal}{\apj} \textbf{\bibinfo{volume}{414}}, \bibinfo{pages}{247}
  (\bibinfo{year}{1993}).

\bibitem[{\citenamefont{{Andersson} and {Kokkotas}}(2001)}]{Andersson01}
\bibinfo{author}{\bibfnamefont{N.}~\bibnamefont{{Andersson}}} \bibnamefont{and}
  \bibinfo{author}{\bibfnamefont{K.~D.} \bibnamefont{{Kokkotas}}},
  \bibinfo{journal}{\ijmpd} \textbf{\bibinfo{volume}{10}}, \bibinfo{pages}{381}
  (\bibinfo{year}{2001}).

\bibitem[{\citenamefont{Ruoff and Kokkotas}(2001)}]{Ruoff01}
\bibinfo{author}{\bibfnamefont{J.}~\bibnamefont{Ruoff}} \bibnamefont{and}
  \bibinfo{author}{\bibfnamefont{K.~D.} \bibnamefont{Kokkotas}},
  \bibinfo{journal}{\mnras} \textbf{\bibinfo{volume}{328}},
  \bibinfo{pages}{678} (\bibinfo{year}{2001}).

\bibitem[{\citenamefont{Gaertig and Kokkotas}(2008)}]{Gaertig08}
\bibinfo{author}{\bibfnamefont{E.}~\bibnamefont{Gaertig}} \bibnamefont{and}
  \bibinfo{author}{\bibfnamefont{K.~D.} \bibnamefont{Kokkotas}},
  \bibinfo{journal}{\prd} \textbf{\bibinfo{volume}{78}},
  \bibinfo{pages}{064063} (\bibinfo{year}{2008}).

\bibitem[{\citenamefont{Gaertig and Kokkotas}(2009)}]{Gaertig09}
\bibinfo{author}{\bibfnamefont{E.}~\bibnamefont{Gaertig}} \bibnamefont{and}
  \bibinfo{author}{\bibfnamefont{K.~D.} \bibnamefont{Kokkotas}},
  \bibinfo{journal}{\prd} \textbf{\bibinfo{volume}{80}},
  \bibinfo{pages}{064026} (\bibinfo{year}{2009}).

\bibitem[{\citenamefont{Kr\"uger et~al.}(2010)\citenamefont{Kr\"uger, Gaertig,
  and Kokkotas}}]{Kruger10}
\bibinfo{author}{\bibfnamefont{C.}~\bibnamefont{Kr\"uger}},
  \bibinfo{author}{\bibfnamefont{E.}~\bibnamefont{Gaertig}}, \bibnamefont{and}
  \bibinfo{author}{\bibfnamefont{K.~D.} \bibnamefont{Kokkotas}},
  \bibinfo{journal}{\prd} \textbf{\bibinfo{volume}{81}},
  \bibinfo{pages}{084019} (\bibinfo{year}{2010}).

\bibitem[{\citenamefont{Kr\"uger}(2009)}]{KrugerMaster}
\bibinfo{author}{\bibfnamefont{C.}~\bibnamefont{Kr\"uger}}, Master's thesis,
  \bibinfo{school}{Eberhard Karls Universit\"at, T\"ubingen}
  (\bibinfo{year}{2009}).

\bibitem[{\citenamefont{Gaertig et~al.}(2011)\citenamefont{Gaertig,
  Glampedakis, Kokkotas, and Zink}}]{Gaertig11}
\bibinfo{author}{\bibfnamefont{E.}~\bibnamefont{Gaertig}},
  \bibinfo{author}{\bibfnamefont{K.}~\bibnamefont{Glampedakis}},
  \bibinfo{author}{\bibfnamefont{K.~D.} \bibnamefont{Kokkotas}},
  \bibnamefont{and} \bibinfo{author}{\bibfnamefont{B.}~\bibnamefont{Zink}},
  \bibinfo{journal}{\prl} \textbf{\bibinfo{volume}{107}},
  \bibinfo{pages}{101102} (\bibinfo{year}{2011}).

\bibitem[{\citenamefont{Passamonti et~al.}(2013)\citenamefont{Passamonti,
  Gaertig, Kokkotas, and Doneva}}]{Passamonti12}
\bibinfo{author}{\bibfnamefont{A.}~\bibnamefont{Passamonti}},
  \bibinfo{author}{\bibfnamefont{E.}~\bibnamefont{Gaertig}},
  \bibinfo{author}{\bibfnamefont{K.}~\bibnamefont{Kokkotas}}, \bibnamefont{and}
  \bibinfo{author}{\bibfnamefont{D.~D.} \bibnamefont{Doneva}},
  \bibinfo{journal}{\prd}  (\bibinfo{year}{2013}).

\bibitem[{\citenamefont{{Yoshida}}(2012)}]{Yoshida12}
\bibinfo{author}{\bibfnamefont{S.}~\bibnamefont{{Yoshida}}},
  \bibinfo{journal}{\prd} \textbf{\bibinfo{volume}{86}}, \bibinfo{eid}{104055}
  (\bibinfo{year}{2012}).

\bibitem[{\citenamefont{Doneva and Yazadjiev}(2012)}]{Doneva12}
\bibinfo{author}{\bibfnamefont{D.~D.} \bibnamefont{Doneva}} \bibnamefont{and}
  \bibinfo{author}{\bibfnamefont{S.~S.} \bibnamefont{Yazadjiev}},
  \bibinfo{journal}{\prd} \textbf{\bibinfo{volume}{85}},
  \bibinfo{pages}{124023} (\bibinfo{year}{2012}).

\bibitem[{\citenamefont{Yazadjiev and Doneva}(2012)}]{Yazadjiev12}
\bibinfo{author}{\bibfnamefont{S.~S.} \bibnamefont{Yazadjiev}}
  \bibnamefont{and} \bibinfo{author}{\bibfnamefont{D.~D.}
  \bibnamefont{Doneva}}, \bibinfo{journal}{\jcap}
  \textbf{\bibinfo{volume}{1203}}, \bibinfo{pages}{037} (\bibinfo{year}{2012}).

\bibitem[{\citenamefont{Sotani and Kokkotas}(2004)}]{Sotani04}
\bibinfo{author}{\bibfnamefont{H.}~\bibnamefont{Sotani}} \bibnamefont{and}
  \bibinfo{author}{\bibfnamefont{K.~D.} \bibnamefont{Kokkotas}},
  \bibinfo{journal}{\prd} \textbf{\bibinfo{volume}{70}},
  \bibinfo{pages}{084026} (\bibinfo{year}{2004}).

\bibitem[{\citenamefont{Sotani}(2009)}]{Sotani09}
\bibinfo{author}{\bibfnamefont{H.}~\bibnamefont{Sotani}},
  \bibinfo{journal}{\prd} \textbf{\bibinfo{volume}{79}},
  \bibinfo{pages}{064033} (\bibinfo{year}{2009}).

\bibitem[{\citenamefont{Kokkotas and Vavoulidis}(2005)}]{Vavoulidis05}
\bibinfo{author}{\bibfnamefont{K.~D.} \bibnamefont{Kokkotas}} \bibnamefont{and}
  \bibinfo{author}{\bibfnamefont{M.}~\bibnamefont{Vavoulidis}},
  \bibinfo{journal}{\jpcs} \textbf{\bibinfo{volume}{8}}, \bibinfo{pages}{71}
  (\bibinfo{year}{2005}).

\bibitem[{\citenamefont{Vavoulidis}(2007)}]{Vavoulidis07}
\bibinfo{author}{\bibfnamefont{M.}~\bibnamefont{Vavoulidis}}, Ph.D. thesis,
  \bibinfo{school}{Aristotle University of Thessaloniki}
  (\bibinfo{year}{2007}).

\bibitem[{\citenamefont{{Cowling}}(1941)}]{Cowling41}
\bibinfo{author}{\bibfnamefont{T.~G.} \bibnamefont{{Cowling}}},
  \bibinfo{journal}{\mnras} \textbf{\bibinfo{volume}{101}},
  \bibinfo{pages}{367} (\bibinfo{year}{1941}).

\bibitem[{\citenamefont{{McDermott} et~al.}(1983)\citenamefont{{McDermott},
  {van Horn}, and {Scholl}}}]{McDermott83}
\bibinfo{author}{\bibfnamefont{P.~N.} \bibnamefont{{McDermott}}},
  \bibinfo{author}{\bibfnamefont{H.~M.} \bibnamefont{{van Horn}}},
  \bibnamefont{and} \bibinfo{author}{\bibfnamefont{J.~F.}
  \bibnamefont{{Scholl}}}, \bibinfo{journal}{\apj}
  \textbf{\bibinfo{volume}{268}}, \bibinfo{pages}{837} (\bibinfo{year}{1983}).

\bibitem[{\citenamefont{{Finn}}(1988)}]{Finn88}
\bibinfo{author}{\bibfnamefont{L.~S.} \bibnamefont{{Finn}}},
  \bibinfo{journal}{\mnras} \textbf{\bibinfo{volume}{232}},
  \bibinfo{pages}{259} (\bibinfo{year}{1988}).

\bibitem[{\citenamefont{{Lindblom} and {Splinter}}(1990)}]{Lindblom90}
\bibinfo{author}{\bibfnamefont{L.}~\bibnamefont{{Lindblom}}} \bibnamefont{and}
  \bibinfo{author}{\bibfnamefont{R.~J.} \bibnamefont{{Splinter}}},
  \bibinfo{journal}{\apj} \textbf{\bibinfo{volume}{348}}, \bibinfo{pages}{198}
  (\bibinfo{year}{1990}).

\bibitem[{\citenamefont{{Yoshida} and {Kojima}}(1997)}]{Yoshida97}
\bibinfo{author}{\bibfnamefont{S.}~\bibnamefont{{Yoshida}}} \bibnamefont{and}
  \bibinfo{author}{\bibfnamefont{Y.}~\bibnamefont{{Kojima}}},
  \bibinfo{journal}{\mnras} \textbf{\bibinfo{volume}{289}},
  \bibinfo{pages}{117} (\bibinfo{year}{1997}).

\bibitem[{\citenamefont{{Thorne}}(1980)}]{Thorne80}
\bibinfo{author}{\bibfnamefont{K.~S.} \bibnamefont{{Thorne}}},
  \bibinfo{journal}{\rmp} \textbf{\bibinfo{volume}{52}}, \bibinfo{pages}{299}
  (\bibinfo{year}{1980}).

\bibitem[{\citenamefont{{Balbinski} and {Schutz}}(1982)}]{Balbinski82}
\bibinfo{author}{\bibfnamefont{E.}~\bibnamefont{{Balbinski}}} \bibnamefont{and}
  \bibinfo{author}{\bibfnamefont{B.~F.} \bibnamefont{{Schutz}}},
  \bibinfo{journal}{\mnras} \textbf{\bibinfo{volume}{200}},
  \bibinfo{pages}{43P} (\bibinfo{year}{1982}).

\bibitem[{\citenamefont{Balbinski et~al.}(1985)\citenamefont{Balbinski,
  Detweiler, Lindblom, and Schutz}}]{Balbinski85}
\bibinfo{author}{\bibfnamefont{E.}~\bibnamefont{Balbinski}},
  \bibinfo{author}{\bibfnamefont{S.~L.} \bibnamefont{Detweiler}},
  \bibinfo{author}{\bibfnamefont{L.}~\bibnamefont{Lindblom}}, \bibnamefont{and}
  \bibinfo{author}{\bibfnamefont{B.}~\bibnamefont{Schutz}},
  \bibinfo{journal}{\mnras} \textbf{\bibinfo{volume}{213}},
  \bibinfo{pages}{553} (\bibinfo{year}{1985}).

\bibitem[{\citenamefont{{Lockitch} and {Friedman}}(1999)}]{Lockitch98}
\bibinfo{author}{\bibfnamefont{K.~H.} \bibnamefont{{Lockitch}}}
  \bibnamefont{and} \bibinfo{author}{\bibfnamefont{J.~L.}
  \bibnamefont{{Friedman}}}, \bibinfo{journal}{\apj}
  \textbf{\bibinfo{volume}{521}}, \bibinfo{pages}{764} (\bibinfo{year}{1999}).

\bibitem[{\citenamefont{{Ipser} and {Lindblom}}(1991)}]{Ipser91}
\bibinfo{author}{\bibfnamefont{J.~R.} \bibnamefont{{Ipser}}} \bibnamefont{and}
  \bibinfo{author}{\bibfnamefont{L.}~\bibnamefont{{Lindblom}}},
  \bibinfo{journal}{\apj} \textbf{\bibinfo{volume}{373}}, \bibinfo{pages}{213}
  (\bibinfo{year}{1991}).

\bibitem[{\citenamefont{Kastaun et~al.}(2010)\citenamefont{Kastaun, Willburger,
  and Kokkotas}}]{Kastaun10}
\bibinfo{author}{\bibfnamefont{W.}~\bibnamefont{Kastaun}},
  \bibinfo{author}{\bibfnamefont{B.}~\bibnamefont{Willburger}},
  \bibnamefont{and} \bibinfo{author}{\bibfnamefont{K.~D.}
  \bibnamefont{Kokkotas}}, \bibinfo{journal}{\prd}
  \textbf{\bibinfo{volume}{82}}, \bibinfo{pages}{104036}
  (\bibinfo{year}{2010}).

\bibitem[{\citenamefont{{Sawyer}}(1989)}]{Sawyer89}
\bibinfo{author}{\bibfnamefont{R.~F.} \bibnamefont{{Sawyer}}},
  \bibinfo{journal}{\prd} \textbf{\bibinfo{volume}{39}}, \bibinfo{pages}{3804}
  (\bibinfo{year}{1989}).

\bibitem[{\citenamefont{{Cutler} and {Lindblom}}(1987)}]{Cutler87}
\bibinfo{author}{\bibfnamefont{C.}~\bibnamefont{{Cutler}}} \bibnamefont{and}
  \bibinfo{author}{\bibfnamefont{L.}~\bibnamefont{{Lindblom}}},
  \bibinfo{journal}{\apj} \textbf{\bibinfo{volume}{314}}, \bibinfo{pages}{234}
  (\bibinfo{year}{1987}).

\bibitem[{\citenamefont{{Alford} et~al.}(2012)\citenamefont{{Alford},
  {Mahmoodifar}, and {Schwenzer}}}]{Alford12}
\bibinfo{author}{\bibfnamefont{M.~G.} \bibnamefont{{Alford}}},
  \bibinfo{author}{\bibfnamefont{S.}~\bibnamefont{{Mahmoodifar}}},
  \bibnamefont{and}
  \bibinfo{author}{\bibfnamefont{K.}~\bibnamefont{{Schwenzer}}},
  \bibinfo{journal}{\prd} \textbf{\bibinfo{volume}{85}}, \bibinfo{eid}{044051}
  (\bibinfo{year}{2012}).

\bibitem[{\citenamefont{Stergioulas and Friedman}(1995)}]{Stergioulas95}
\bibinfo{author}{\bibfnamefont{N.}~\bibnamefont{Stergioulas}} \bibnamefont{and}
  \bibinfo{author}{\bibfnamefont{J.}~\bibnamefont{Friedman}},
  \bibinfo{journal}{\apj} \textbf{\bibinfo{volume}{444}}, \bibinfo{pages}{306}
  (\bibinfo{year}{1995}).

\bibitem[{\citenamefont{Nozawa et~al.}(1998)\citenamefont{Nozawa, Stergioulas,
  Gourgoulhon, and Eriguchi}}]{Nozawa98}
\bibinfo{author}{\bibfnamefont{T.}~\bibnamefont{Nozawa}},
  \bibinfo{author}{\bibfnamefont{N.}~\bibnamefont{Stergioulas}},
  \bibinfo{author}{\bibfnamefont{E.}~\bibnamefont{Gourgoulhon}},
  \bibnamefont{and} \bibinfo{author}{\bibfnamefont{Y.}~\bibnamefont{Eriguchi}},
  \bibinfo{journal}{\aaps} \textbf{\bibinfo{volume}{132}}, \bibinfo{pages}{431}
  (\bibinfo{year}{1998}).

\bibitem[{\citenamefont{{Demorest} et~al.}(2010)\citenamefont{{Demorest},
  {Pennucci}, {Ransom}, {Roberts}, and {Hessels}}}]{Demorest10}
\bibinfo{author}{\bibfnamefont{P.~B.} \bibnamefont{{Demorest}}},
  \bibinfo{author}{\bibfnamefont{T.}~\bibnamefont{{Pennucci}}},
  \bibinfo{author}{\bibfnamefont{S.~M.} \bibnamefont{{Ransom}}},
  \bibinfo{author}{\bibfnamefont{M.~S.~E.} \bibnamefont{{Roberts}}},
  \bibnamefont{and} \bibinfo{author}{\bibfnamefont{J.~W.~T.}
  \bibnamefont{{Hessels}}}, \bibinfo{journal}{\nat}
  \textbf{\bibinfo{volume}{467}}, \bibinfo{pages}{1081} (\bibinfo{year}{2010}).

\bibitem[{\citenamefont{{Manousakis} et~al.}(2012)\citenamefont{{Manousakis},
  {Walter}, and {Blondin}}}]{Manousakis12}
\bibinfo{author}{\bibfnamefont{A.}~\bibnamefont{{Manousakis}}},
  \bibinfo{author}{\bibfnamefont{R.}~\bibnamefont{{Walter}}}, \bibnamefont{and}
  \bibinfo{author}{\bibfnamefont{J.~M.} \bibnamefont{{Blondin}}},
  \bibinfo{journal}{\aap} \textbf{\bibinfo{volume}{547}}, \bibinfo{eid}{A20}
  (\bibinfo{year}{2012}).

\bibitem[{\citenamefont{Antoniadis et~al.}(2013)\citenamefont{Antoniadis,
  Freire, Wex, Tauris, Lynch et~al.}}]{Antoniadis13}
\bibinfo{author}{\bibfnamefont{J.}~\bibnamefont{Antoniadis}},
  \bibinfo{author}{\bibfnamefont{P.~C.} \bibnamefont{Freire}},
  \bibinfo{author}{\bibfnamefont{N.}~\bibnamefont{Wex}},
  \bibinfo{author}{\bibfnamefont{T.~M.} \bibnamefont{Tauris}},
  \bibinfo{author}{\bibfnamefont{R.~S.} \bibnamefont{Lynch}},
  \bibnamefont{et~al.}, \bibinfo{journal}{Science}
  \textbf{\bibinfo{volume}{340}}, \bibinfo{pages}{6131} (\bibinfo{year}{2013}).

\bibitem[{\citenamefont{{Lattimer}}(2012)}]{Lattimer12}
\bibinfo{author}{\bibfnamefont{J.}~\bibnamefont{{Lattimer}}},
  \bibinfo{journal}{Annu. Rev. Nucl. Part. Sci.} \textbf{\bibinfo{volume}{62}},
  \bibinfo{pages}{485} (\bibinfo{year}{2012}).

\bibitem[{\citenamefont{{Lorenz} et~al.}(1993)\citenamefont{{Lorenz},
  {Ravenhall}, and {Pethick}}}]{FPS}
\bibinfo{author}{\bibfnamefont{C.~P.} \bibnamefont{{Lorenz}}},
  \bibinfo{author}{\bibfnamefont{D.~G.} \bibnamefont{{Ravenhall}}},
  \bibnamefont{and} \bibinfo{author}{\bibfnamefont{C.~J.}
  \bibnamefont{{Pethick}}}, \bibinfo{journal}{\prl}
  \textbf{\bibinfo{volume}{70}}, \bibinfo{pages}{379} (\bibinfo{year}{1993}).

\bibitem[{\citenamefont{{Friedman} and
  {Pandharipande}}(1981)}]{Pandharipande81}
\bibinfo{author}{\bibfnamefont{B.}~\bibnamefont{{Friedman}}} \bibnamefont{and}
  \bibinfo{author}{\bibfnamefont{V.~R.} \bibnamefont{{Pandharipande}}},
  \bibinfo{journal}{\npa} \textbf{\bibinfo{volume}{361}}, \bibinfo{pages}{502}
  (\bibinfo{year}{1981}).

\bibitem[{\citenamefont{{Wiringa} et~al.}(1988)\citenamefont{{Wiringa}, {Fiks},
  and {Fabrocini}}}]{WFF}
\bibinfo{author}{\bibfnamefont{R.~B.} \bibnamefont{{Wiringa}}},
  \bibinfo{author}{\bibfnamefont{V.}~\bibnamefont{{Fiks}}}, \bibnamefont{and}
  \bibinfo{author}{\bibfnamefont{A.}~\bibnamefont{{Fabrocini}}},
  \bibinfo{journal}{\prc} \textbf{\bibinfo{volume}{38}}, \bibinfo{pages}{1010}
  (\bibinfo{year}{1988}).

\bibitem[{\citenamefont{{Negele} and {Vautherin}}(1973)}]{Negele73}
\bibinfo{author}{\bibfnamefont{J.~W.} \bibnamefont{{Negele}}} \bibnamefont{and}
  \bibinfo{author}{\bibfnamefont{D.}~\bibnamefont{{Vautherin}}},
  \bibinfo{journal}{\npa} \textbf{\bibinfo{volume}{207}}, \bibinfo{pages}{298}
  (\bibinfo{year}{1973}).

\bibitem[{\citenamefont{{Pandharipande}}(1971)}]{Pandharipande71}
\bibinfo{author}{\bibfnamefont{V.~R.} \bibnamefont{{Pandharipande}}},
  \bibinfo{journal}{\npa} \textbf{\bibinfo{volume}{174}}, \bibinfo{pages}{641}
  (\bibinfo{year}{1971}).

\bibitem[{\citenamefont{{Arnett} and {Bowers}}(1977)}]{A}
\bibinfo{author}{\bibfnamefont{W.~D.} \bibnamefont{{Arnett}}} \bibnamefont{and}
  \bibinfo{author}{\bibfnamefont{R.~L.} \bibnamefont{{Bowers}}},
  \bibinfo{journal}{\apjs} \textbf{\bibinfo{volume}{33}}, \bibinfo{pages}{415}
  (\bibinfo{year}{1977}).

\bibitem[{\citenamefont{{Akmal} et~al.}(1998)\citenamefont{{Akmal},
  {Pandharipande}, and {Ravenhall}}}]{AkmalPR}
\bibinfo{author}{\bibfnamefont{A.}~\bibnamefont{{Akmal}}},
  \bibinfo{author}{\bibfnamefont{V.~R.} \bibnamefont{{Pandharipande}}},
  \bibnamefont{and} \bibinfo{author}{\bibfnamefont{D.~G.}
  \bibnamefont{{Ravenhall}}}, \bibinfo{journal}{\prc}
  \textbf{\bibinfo{volume}{58}}, \bibinfo{pages}{1804} (\bibinfo{year}{1998}).

\bibitem[{\citenamefont{{Douchin} and {Haensel}}(2001)}]{Douchin01}
\bibinfo{author}{\bibfnamefont{F.}~\bibnamefont{{Douchin}}} \bibnamefont{and}
  \bibinfo{author}{\bibfnamefont{P.}~\bibnamefont{{Haensel}}},
  \bibinfo{journal}{A\&A} \textbf{\bibinfo{volume}{380}}, \bibinfo{pages}{151}
  (\bibinfo{year}{2001}).

\bibitem[{\citenamefont{Friedman et~al.}(1989)\citenamefont{Friedman, Ipser,
  and Parker}}]{Friedman89}
\bibinfo{author}{\bibfnamefont{J.~L.} \bibnamefont{Friedman}},
  \bibinfo{author}{\bibfnamefont{J.~R.} \bibnamefont{Ipser}}, \bibnamefont{and}
  \bibinfo{author}{\bibfnamefont{L.}~\bibnamefont{Parker}},
  \bibinfo{journal}{\prl} \textbf{\bibinfo{volume}{62}}, \bibinfo{pages}{3015}
  (\bibinfo{year}{1989}).

\bibitem[{\citenamefont{{Haensel} and {Zdunik}}(1989)}]{Haensel89}
\bibinfo{author}{\bibfnamefont{P.}~\bibnamefont{{Haensel}}} \bibnamefont{and}
  \bibinfo{author}{\bibfnamefont{J.~L.} \bibnamefont{{Zdunik}}},
  \bibinfo{journal}{\nat} \textbf{\bibinfo{volume}{340}}, \bibinfo{pages}{617}
  (\bibinfo{year}{1989}).

\bibitem[{\citenamefont{{Lasota} et~al.}(1996)\citenamefont{{Lasota},
  {Haensel}, and {Abramowicz}}}]{Lasota96}
\bibinfo{author}{\bibfnamefont{J.-P.} \bibnamefont{{Lasota}}},
  \bibinfo{author}{\bibfnamefont{P.}~\bibnamefont{{Haensel}}},
  \bibnamefont{and} \bibinfo{author}{\bibfnamefont{M.~A.}
  \bibnamefont{{Abramowicz}}}, \bibinfo{journal}{\apj}
  \textbf{\bibinfo{volume}{456}}, \bibinfo{pages}{300} (\bibinfo{year}{1996}).

\bibitem[{\citenamefont{{Andersson} and {Kokkotas}}(1998)}]{Andersson98c}
\bibinfo{author}{\bibfnamefont{N.}~\bibnamefont{{Andersson}}} \bibnamefont{and}
  \bibinfo{author}{\bibfnamefont{K.~D.} \bibnamefont{{Kokkotas}}},
  \bibinfo{journal}{\mnras} \textbf{\bibinfo{volume}{297}},
  \bibinfo{pages}{493} (\bibinfo{year}{1998}).

\bibitem[{\citenamefont{{Detweiler}}(1975)}]{Detweiler75}
\bibinfo{author}{\bibfnamefont{S.~L.} \bibnamefont{{Detweiler}}},
  \bibinfo{journal}{\apj} \textbf{\bibinfo{volume}{197}}, \bibinfo{pages}{203}
  (\bibinfo{year}{1975}).

\bibitem[{\citenamefont{{Andersson}}(1998)}]{Andersson98b}
\bibinfo{author}{\bibfnamefont{N.}~\bibnamefont{{Andersson}}},
  \bibinfo{journal}{\apj} \textbf{\bibinfo{volume}{502}}, \bibinfo{pages}{708}
  (\bibinfo{year}{1998}).

\bibitem[{\citenamefont{{Friedman} and {Morsink}}(1998)}]{Friedman98}
\bibinfo{author}{\bibfnamefont{J.~L.} \bibnamefont{{Friedman}}}
  \bibnamefont{and} \bibinfo{author}{\bibfnamefont{S.~M.}
  \bibnamefont{{Morsink}}}, \bibinfo{journal}{\apj}
  \textbf{\bibinfo{volume}{502}}, \bibinfo{pages}{714} (\bibinfo{year}{1998}).

\bibitem[{\citenamefont{{Kokkotas} and {Stergioulas}}(1999)}]{Kokkotas99b}
\bibinfo{author}{\bibfnamefont{K.~D.} \bibnamefont{{Kokkotas}}}
  \bibnamefont{and}
  \bibinfo{author}{\bibfnamefont{N.}~\bibnamefont{{Stergioulas}}},
  \bibinfo{journal}{\aap} \textbf{\bibinfo{volume}{341}}, \bibinfo{pages}{110}
  (\bibinfo{year}{1999}).

\bibitem[{\citenamefont{{Bondarescu} et~al.}(2007)\citenamefont{{Bondarescu},
  {Teukolsky}, and {Wasserman}}}]{Bondarescu07}
\bibinfo{author}{\bibfnamefont{R.}~\bibnamefont{{Bondarescu}}},
  \bibinfo{author}{\bibfnamefont{S.~A.} \bibnamefont{{Teukolsky}}},
  \bibnamefont{and}
  \bibinfo{author}{\bibfnamefont{I.}~\bibnamefont{{Wasserman}}},
  \bibinfo{journal}{\prd} \textbf{\bibinfo{volume}{76}}, \bibinfo{eid}{064019}
  (\bibinfo{year}{2007}).

\bibitem[{\citenamefont{{Schenk} et~al.}(2002)\citenamefont{{Schenk}, {Arras},
  {Flanagan}, {Teukolsky}, and {Wasserman}}}]{Schenk02}
\bibinfo{author}{\bibfnamefont{A.~K.} \bibnamefont{{Schenk}}},
  \bibinfo{author}{\bibfnamefont{P.}~\bibnamefont{{Arras}}},
  \bibinfo{author}{\bibfnamefont{{\'E}.~{\'E}.} \bibnamefont{{Flanagan}}},
  \bibinfo{author}{\bibfnamefont{S.~A.} \bibnamefont{{Teukolsky}}},
  \bibnamefont{and}
  \bibinfo{author}{\bibfnamefont{I.}~\bibnamefont{{Wasserman}}},
  \bibinfo{journal}{\prd} \textbf{\bibinfo{volume}{65}}, \bibinfo{eid}{024001}
  (\bibinfo{year}{2002}).

\end{thebibliography}

\end{document}